\documentstyle[aps,amsfonts,eqsecnum,multicol,prb]{revtex}
\def\rmi{{\rm i}}
\def\rme{{\rm e}}
\def\rmd{{\rm d}}
\def\mbr{{\bbox{r}}}
\def\mbq{{\bbox{q}}}
\def\Re{\hbox{\rm Re}}
\def\Im{\hbox{\rm Im}}
\def\Tr{\hbox{\rm Tr}}

\def\calE{{\cal{E}}}
\def\calK{{\cal K}}

\def\calO{{\cal O}}
\def\calP{{\cal P}}
\def\calQ{{\cal Q}}

\def\calR{{\cal R}} 
\def\calT{{\cal T}}
\def\sfD{{\sf D}}
\def\sfN{{\sf N}}
\def\sfQ{{\sf Q}}
\begin{document}
\preprint{}%\today }

\draft

%\tighten

\title{Design of semiconductor heterostructures with preset electron
reflectance by inverse scattering techniques}
\author{Daniel Bessis,$^{1,2}$ and G. Andrei Mezincescu$^{3}$
\protect{\footnote{E-mail mezin@alpha1.infim.ro}}}
\address{$^1$ CTSPS, Clark-Atlanta University, Atlanta, GA 30314\\
$^2$ Service de Physique Th\'eorique, C.E. Saclay,
F-91191 Gif-sur-Yvette Cedex, France\\ 
$^3$ Institutul Na\c tional de Fizica Materialelor, C.P. MG-7, 
R-76900 Bucure\c sti -- M\u agurele, Rom\^ania}

\maketitle

\begin{abstract}
We present the application of the inverse scattering method
to the design of semiconductor heterostructures having a
preset dependence of the (conduction) electrons' reflectance
on the energy. The electron dynamics are described by either
the effective mass Schr\"odinger, or by the (variable
mass) BenDaniel and Duke equations. The problem of phase (re)construction
for the complex transmission and reflection coefficients is
solved by a combination of Pad\'e approximant techniques, obtaining
reference solutions with simple analytic properties. 
Reflectance-preserving transformations allow bound state and reflection
resonance management. The inverse scattering problem for the Schr\"odinger
equation is solved using an algebraic approach due to Sabatier.
This solution can be mapped unitarily onto a family of BenDaniel and
Duke type equations. The boundary value problem for the
nonlinear equation which determines the mapping is discussed in some detail. 
The chemical concentration profile of heterostructures whose self consistent
potential yields the desired reflectance is solved completely in the
case of Schr\"odinger dynamics and approximately for
Ben-Daniel and Duke dynamics. The Appendix contains a brief digest of results 
from scattering and inverse scattering theory for the one-dimensional Schr\"odinger 
equation which are used in the paper.
\end{abstract}

%{\bf Draft: }

\begin{multicols}{2}

\narrowtext

{{\small \tableofcontents}}

\section{Introduction}\label{i}

A semiconductor heterostructure can be modelled by a system of 
equations describing (with a certain degree of completeness and precision)
the state of the system. The equations depend on a set of
structural and compositional data (SCD). Essentially, these are 
the spatial dependence of the chemical composition (including dopant 
profiles), the applied external fields, etc.                                                          
The system's behavior (response) is described by functional data (FD), 
such as the electric or thermal conductance, the energy dependence of the
electron transmittance, the wavelength dependence of the optical absorption  
coefficient, etc.
The FD can be computed using the solution of the equations, and are thus
functionals of the SCD. 

To design a heterostructure for a certain application is to find a set
of SCD, which is physically (and technologically) achievable, such that the
values of a chosen subset of FD will be within desirable ranges. 
The designer solves thus an inverse problem: inverting the functional 
dependence of the FD on the SCD.
This problem is rather ill-posed. The desired ranges of FP may be unachievable.
Generally speaking, even if a certain desired set of FD values is achievable, the
set of SCD which achieves it is not unique. This absence of uniqueness is
not bad in itself. If several solutions {\em can be obtained}, then
one may further optimize the design in terms of other properties which
were not included in the original specifications. The difficulty is
mathematical. Inverting a one-to-one functional dependence can be a
formidable task, which is further aggravated by the lack of uniqueness.

A brute-force approach to the problem is always possible: computing the 
FD for a set of achievable SCD (ideally all) and selecting the best. 
In practice, brute force optimization is restricted to rather small sets 
of parameters describing the FD. This is mostly due the fact that any 
conceivable penalty function will be non-convex. Its graph will have a rather 
complex multi-valley shape. The search algorithm is forced to do a thorough 
investigation of this landscape and will eventually fail through run time 
limitations. The intelligent designer will partially avoid such restrictions
by noting trends, trying to break the design into combining manageable blocks
and locally improving promising configurations.

Thus, cases when the solution of the inverse problem is realizable by methods 
which are less costly than the brute force approach can be rather useful. 
Even if one has to simplify somewhat the physical model, precious insights 
on new promising configurations can be obtained.

In this paper we will try to review the possible applications of inverse 
scattering techniques to some aspects of heterostructure design.
Our physical model for the electron states in the heterostructure will be the 
(one band) effective mass approximation for the envelope function
of the (conduction band) electrons. 

The simplest approach is to assume a constant effective mass. Then, 
the envelopes of electron states satisfy an effective Schr\"odinger 
equation\cite{Bastard}. 
Choosing the $z$-axis along the growth direction, 
\begin{equation}
\rmi\hbar\frac{\partial}{\partial t}\Psi(\mbr,t) =
-\frac{\hbar^2}{2m_e}\Delta\Psi (\mbr,t)+U(z)\Psi(\mbr,t).
\label{S1}\end{equation}

Here, $m_e$ is the (conduction band) electron's effective mass. 
The potential is
\begin{equation}
U(z)=\calE(z)+U_{ext}(z)+\Phi_{sc}(z),
\label{S2}\end{equation}
where $\calE(z)=\calE_{cond}[c(z)]$ is the (conduction) band offset 
(assumed to depend only on the local chemical composition, $c(z)$); 
$U_{ext}(z)$ is the (possibly equal to zero) external applied potential 
and we lumped in $\Phi_{sc}(z)$ the potential of the ionized (donor) 
impurities (dopants) and terms which will make the full potential 
$U(z)$ self-consistent. Various models can be considered for $\Phi_{sc}(z)$:  
a Hartree self consistent potential, exchange-correlation corrections can 
be incorporated. The only requirement is that $\Phi_{sc}(z)$ has to be an 
{\it explicitly defined} functional of the full potential $U(z)$.

The next step is to take into account the spatial dependence of the 
effective mass. The Schr\"odinger equation with constant effective mass (SE) 
for the envelope function, (\ref{S1}), is replaced by the BenDaniel and 
Duke\cite{BDD66} equation (BDD):
\begin{equation}
\rmi\hbar\frac{\partial}{\partial t}\Psi(\mbr,t) =
-\bbox{\nabla}\frac{\hbar^2}{2m(z)}\bbox{\nabla}\Psi (\mbr,t)+U(z)\Psi(\mbr,t).
\label{S3}\end{equation}
Here the effective mass of the conduction band electrons 
$m(z)=m_{cond}[c(z)]$ is assumed to depend only on the local chemical 
composition $c(z)$. The self-consistent potential is given again by 
(\ref{S2}). The  $\Phi_{sc}(z)$ term is now a functional of both $m(z)$
and $U(z)$.

In the following we will consider only stationary states of the equations 
(\ref{S1}) and (\ref{S3}). Furthermore, since the potential depends only on 
the coordinate along the growth direction, $z$, the motion in the 
perpendicular plane is free. Setting 
\begin{equation}
\Psi(\mbr,t)=\psi(z)\rme^{\rmi\left(\mbq_\perp\mbr_\perp-Et/\hbar\right)},
\label{none}\end{equation}
in the Schr\"odinger (\ref{S1}) and BenDaniel and Duke (\ref{S3}) equations, 
where $E$ is the energy, 
$\mbr_\perp=(x,y,0)$ and $\mbq_\perp=(q_1,q_2,0)$ are, respectively, 
the coordinates and components of the quasi-momentum in the directions 
perpendicular to the growth axis, we obtain the
one-dimensional Schr\"odinger, 
\begin{equation}
\psi^{\prime\prime}(z)
+\left[k^2-\mbq^2_\perp-V(z)\right]\psi(z) =0,
\label{0-s1}\end{equation}
and BDD 
\begin{equation}
\left[\frac{m_\infty\psi^{\prime}(z)}{m(z)}\right]^\prime
+\left[k^2-\frac{m_\infty\mbq^2_\perp}{m(z)}-V(z)\right]\psi(z) =0,
\label{0-bdd1}\end{equation}
equations for $\psi(z)$. Here, and in the following, we use the 
notation $m_\infty$ for the electron effective mass in the embedding 
material: $m_\infty=m_e$ for SE and $m_\infty=m(\pm\infty)$ for BDD. 
We introduce the notations
\begin{equation}
k=\frac{\sqrt{2m_\infty E}}{\hbar};~~V(z)=\frac{2m_\infty U(z)}{\hbar^2}.
\label{ksq}\end{equation}
Throughout this paper, the square root function is defined with non negative 
imaginary part: $\Im\left(\sqrt{E}\right)\ge 0$. For real positive $E$ 
in (\ref{ksq}), $k>0$. The prime will often be used for derivatives.

The inverse spectral theory for the one-dimensional 
Schr\"odinger equation\cite{PT,ZS} has been successfully applied to some 
optimization problems for the bound states  in semiconductor
quantum wells\cite{ZCh97,MI96,MI97,TMI97,TMI98}. 
Inverse scattering theory\cite{CS} for the one-dimensional Schr\"odinger 
equation, (\ref{0-s1}) with $\mbq_\perp=0$, shows how one can recover the 
potential in (\ref{0-s1}) from the knowledge of the scattering data:
the complex transmission and reflexion to the right/left 
coefficients $\{T(k),\,R_\pm(k)\}$ for all real values of the wave-number $k$.  
Widely used in electric circuit modelling\cite{CS,RT95},
it has been recently applied for designing heterostructure Bloch wave 
filters\cite{BMMV97}.

We want to solve the following problem: let the electron 
dynamics be given by either the SE, (\ref{0-s1}), or by the BDD equation, 
(\ref{0-bdd1}), with the self-consistent potential (\ref{S2}). 
Find chemical composition and dopant profiles, going to constant limits
at infinity, 
such that the heterostructure defined by these data has a given energy 
dependence of the electron reflectance at $\mbq_\perp=0$ and a given
operating temperature:
\begin{equation}
\calR(E)=\left\vert R_\pm(k)\right\vert^2. %
\label{refl}\end{equation}
The zero of the energy scale is chosen at the conduction band minimum 
for the asymptotic composition at infinity.  

In the SE case, the mini-bands are parabolic in $\mbq_\perp$ and (\ref{refl}) will 
hold for $\mbq_\perp\ne 0$ with $E$ changed to $E+\hbar^2\mbq_\perp^2/2m_e$.
This is no longer true in the case of the BDD equation (\ref{0-bdd1}).
For sufficiently small $|\mbq_\perp|$,
the mini-bands will be approximately parabolic only as long as the 
$\frac{m_e\mbq^2_\perp}{m(z)}$ term in the effective potential in (\ref{0-bdd1}) 
can be treated as a first order perturbation.
One could also select a nonzero value of $\mbq^2$ at which (\ref{refl}) is valid,
such as the one corresponding to the transverse thermal energy at the desired 
operating temperature. 

Since only the energy dependence of the reflectance is given,
the first step in solving our problem is to find the sets of
scattering data (SD) which are compatible with $\calR(E)$, {\it i.e.}
find the phases of the scattering data.
In section \ref{pha} we show how to construct SD which correspond to real 
valued potentials $V(z)$ with exponential decay at infinity from $\calR(E)$. 
Physically, one might expect that the potential is determined by the 
its bound state energies and its resonances. 
The reflectance $\calR(E)$ embodies only information on the transmission 
resonances: sharp minima of the reflectance. 
There is another type of resonance, the reflection resonances, 
analogous to the resonances that occur in three-dimensional potential 
scattering on a spherically symmetric potential. These are sharp {\it phase} 
variations of the reflection coefficients.  
The information on reflection resonances and on the bound states 
is not apparent in the reflectance.

The transformations of the SD, which do not change
the reflectance, will be discussed. Using these transformations,
we will define reference solutions for the recovery of the SD from the reflectance.
The reference solutions have no bound states and simpler analytic properties. 
Combinations of reflectance-preserving transformations can then be used to obtain 
the SD of other solutions to the phase reconstruction problem from the 
reference solutions, by dressing them up with bound states and reflection 
resonances. 

We will use Pad\'e approximation methods\cite{baker} to represent the 
scattering data and to find parameterizations for a large class of solutions, 
corresponding to potentials which tend exponentially to zero at infinity.
We will find that on this type of input data, the (re)construction process
amounts essentially to finding the roots of some polynomials and grouping
them into subsets.

In section \ref{rat} we present a simple and efficient algorithm for solving 
the inverse scattering problem for scattering data in the form obtained in 
section \ref{pha}.  We will use results due to Kay, Moses and 
Sabatier\cite{K55,KM56a,KM56b,K60,S83} for the inverse problem with rational 
coefficients\cite{CS}.

In section \ref{vmm} we present the variable mass unitary mapping
of the SE to the BDD equation\cite{BBM95,BBM96,BMMV97}. We formulate
the boundary value problem which must be solved for determining the 
coordinate transformation which defines the mapping, given the material
relation between the effective mass and the band offset. This ill-conditioned 
problem can be solved by a a shooting method. In the case when the relation between
the mass and the offset is linear, the solution takes a simpler form. 
We also give an efficient perturbative method for solving the mapping equation.

In section \ref{v} we show that obtaining chemical composition
profiles and self-consistent potentials for them in the SE in the
inverse scattering approach is simpler than obtaining the self-consistent
potential for a given chemical composition profile. We
also discuss the functional equation which must be solved
for obtaining the chemical (effective mass) profile corresponding
to self-consistent potentials in the BenDaniel and Duke's equation.

For the reader's convenience, in Appendix \ref{appA} we give a brief outline
of results from scattering and inverse scattering theory, which are needed 
and often referred to in the main body of the paper.
%Finally, in
%Appendix \ref{appB} we give an outline on the Darboux transformation method 
%for managing the problems of bound states and reflection resonances.

\section{Phase reconstruction}\label{pha}

Inverse scattering theory for the one-dimensional Schr\"odinger equation, 
(\ref{0-s1}) with $\mbq_\perp=0$, 
\begin{equation}
\psi^{\prime\prime}(z)
+\left[k^2-V(z)\right]\psi(z) =0,
\label{2s} \end{equation}
on which we give a primer in Appendix \ref{appA}, shows that 
one can recover a fast decaying and piecewise continuous potential in 
(\ref{2s}) from the knowledge of the scattering data (SD): the complex 
transmission and reflexion to the right/left coefficients 
$\{T(k),\,R_\pm(k)\}$ 
for all real values of the wave-number $k$ {\it if there are no bound states}.  
If bound states are present, knowledge of the SD is not sufficient for unique
recovery of the potential. If the number of bound states is exactly $n$,
then a $n$-parameter family of potentials gives exactly the same scattering
data. %\cite{parratio}

The SD are completely determined by one of the reflexion coefficients and
the values of the energies of the bound states. 
The phase of $T(k)$ can be obtained from a logarithmic dispersion relation.
(see {\it e.g.} the book by Chadan and Sabatier\cite{CS}, XVII.1.5). The
other reflection coefficient can be obtained from (\ref{7a}).

In this section we assume that the reflectance, $\calR(E)$, is known on the
positive energy half-axis, $E>0$. We want to construct sets of SD which
satisfy (\ref{refl}).           
Since the values of the SD for scattering by short-range and piecewise
continuous potentials must satisfy the constraints\cite{CS,DT79} which are
enumerated at the end of Appendix \ref{i-M}, the function $\calR(E)$ cannot
be arbitrary. It must be non-negative and smaller than unity, with the
exception of $\calR(0)$, which is generically\cite{calR0} equal to 1. 
If the potential is piecewise continuous, $\calR(E)$ must go to zero no 
slower than $E^{-2}$ for large values of $E$. 

The transmittance $\calT(E)=\left\vert T(k)\right\vert^2$, where $T(k)$ 
is the complex transmission coefficient, is readily recovered from 
$\calR(E)+\calT(E)=1$. Thus, we know the absolute values of the scattering 
data and we need the phases. 

\subsection{Reflectance-preserving transformations}\label{pha-1}

The problem of finding the phases of the scattering coefficients knowing only their
absolute values on the real axis is underdetermined and has an infinite
number  of solutions. 
Before considering the phase (re)construction problem, we will introduce
two types of transformations which modify the phases of the 
scattering data without changing the reflectance. 

Let the set of scattering data 
\begin{equation}
\{T(k),\,R_+(k)\,R_-(k)\},\label{sd}
\end{equation}
be a solution  of the phase reconstruction problem, {\it i.e.} the 
scattering coefficients satisfy (\ref{refl}) and the conditions enumerated
at the end of Appendix \ref{i-M}. Then, as mentioned above,  if the SD
(\ref{sd}), has no bound states, there is an unique solution to the inverse 
scattering problem: a potential $V(z)$ in (\ref{2s}) such that the SD 
calculated for this equation coincide with (\ref{sd}). If $T(k)$ has $n$ 
simple imaginary poles in the upper half plane, that is $n$ bound states, 
then a $n$-parameter family of potentials can be constructed, such that the
SD of each potential coincides with (\ref{sd}).
 
Let now $\lambda>0$ be a positive number, such that $k=\rmi\lambda$ is not 
a pole of $T(k)$, {\it i.e.} that $E=-\hbar^2\lambda^2/2m_e$ is not a bound 
state of (\ref{sd}). 
Define a new set of SD by the transformation:
\begin{eqnarray}  %^{(\rmi\lambda)}
{T}(k)&\to&\frac{(k+\rmi\lambda)}{(k-\rmi\lambda)}T(k),
\label{lam1}\\
{R}_\pm(k)&\to&\frac{(\rmi\lambda+k)}{(\rmi\lambda-k)}R_\pm(k),
\label{lam3}     
\end{eqnarray}
The new set of SD will also satisfy the conditions set out at the end of 
Appendix \ref{i-M}, so that a $(n+1)$-parameter family of potentials can 
be constructed with each  potential having the SD (\ref{lam1}-\ref{lam3}).

The transformation (\ref{lam1}-\ref{lam3}) can also remove bound states. 
If the initial SD, (\ref{sd}), have a bound state for $E=-\hbar^2\lambda^2/2m_e$, 
then the transformation (\ref{lam1}-\ref{lam3}) 
with $\lambda$ changed into $-\lambda$ in the right-hand sides of the equations, 
transforms that bound state into an {\em anti-bound} state -- an imaginary pole
of $T(k)$ in $\Im(k)<0$, leaving all the others in place.
 
Let $\zeta$ be an arbitrary complex number with nonzero real and imaginary 
parts. Then,  we can define a second type of reflectance preserving 
transformation of the SD (\ref{sd}):
\begin{eqnarray}%^{(\zeta)} \tilde
{T}(k)&\to&T(k),\label{zet1}\\
{R}_-(k)&\to&\frac{(k-\zeta)(k+\zeta^*)}{(k-\zeta^*)(k+\zeta)}R_-(k),\\
{R}_+(k)&\to&\frac{(k-\zeta^*)(k+\zeta)}{(k-\zeta)(k+\zeta^*)}R_+(k).
\label{zet3} \end{eqnarray}
Here and in the following, we use the notation $^*$ for complex conjugation.
The transformed SD, (\ref{zet1}-\ref{zet3}),  have the same reflectance and 
bound states as (\ref{sd}) and satisfy the conditions enumerated at the end 
of Appendix \ref{i-M}. 
Using the inverse scattering method, one can construct from the SD, 
(\ref{zet1}-\ref{zet3}), a new $n$-parameter family of piecewise continuous 
potentials which goes to zero at infinity.

The transformation (\ref{zet1}-\ref{zet3}) has a simple interpretation. 
Assume that the imaginary part of $\zeta$ is much smaller than its  
real part, $|\Im(\zeta)|\ll|Re(\zeta)|$, and the initial reflection 
coefficients, $R_\pm(k)$, are slowly varying on the scale 
$|\Im(\zeta)|$ near $k=\pm\Re(\zeta)$. 
Then, the new scattering data (\ref{zet1}-\ref{zet3}) have a
{\it reflection resonance} of width $|\Im(\zeta)|$ at $k=\pm\Re(\zeta)$. 
Indeed, the phases of the new reflexion coefficients vary by $\pm 2\pi$ in a 
small interval of width $2|\Im(\zeta)|$ centered on $k=\pm\Re(\zeta)$.

A third type of transformation adds purely imaginary resonances
to the reflection coefficients leaving the transmission coefficient unchanged:
\begin{eqnarray}%^{(\rmi\lambda)} \tilde
{T}(k)&\to& T(k), \label{la1}\\
{R}_+(k)&\to&\frac{(\rmi\lambda+k)}{(\rmi\lambda-k)}R_+(k),\\
{R}_-(k)&\to&\frac{(\rmi\lambda-k)}{(\rmi\lambda+k)}R_-(k),
\label{la3}     
\end{eqnarray}
where $\lambda$ is an arbitrary real number. The new SD will also satisfy the 
conditions enumerated at the end of Appendix \ref{i-M}. The reconstruction
of the potential is done exactly as for the second type of transformation.

A sequence of transformations of the second and third type can be written as:
\begin{eqnarray}
T(k)&\to& T(k), \label{lp1}\\
R_-(k)&\to&\frac{S_N(k)}{S_N(-k)}R_-(k),\\
R_+(k)&\to&\frac{S_N(-k)}{S_N(k)}R_+(k), \label{lp3}     
\end{eqnarray}
where $S_N(k)$ is any polynomial whose zeros are invariant with respect
to reflection through the imaginary axis, {\it i.e.} if $\zeta$ is a zero, then
$-\zeta^*$ is also a zero of the polynomial. Such polynomials, normalized by
the condition $S_N(0)=1$, satisfy $S_N(k)=[S_N(-k)]^*$ for real $k$.
Let us recall that an arbitrary polynomial of degree $n$, $\Pi_n(x)$, 
with $\Pi_n(0)=1$, can be expressed through its zeros:
\begin{equation}
\Pi_n(x)=\prod_{i=1}^n\left(1-\frac{x}{x_i}\right).
\label{ro-r}\end{equation}
Here $x_i, ~i=1,\ldots,n$ are the zeros of $\Pi_n(x)$ (including
multiple ones according to their algebraic multiplicity).
Thus, $S_N(k)$ is completely determined.

%In Appendix \ref{appB} we will show how one can define Darboux
%transformations of the one-dimensional Schr\"odinger equation (\ref{2s}).
%The reflectance-preserving transformations (\ref{lam1} - \ref{zet3}) can be
%implemented as simple first- or second-order differential transformations of
%the solutions of (\ref{2s}) and its potential.  

\subsection{The reference solutions}\label{pha-2}

Let us assume that we have found a solution of the phase reconstruction
problem. Generically, it will have some bound states and reflection
resonances. Using suitably chosen transformations of type 
(\ref{lam1}-\ref{lam3}) one can obtain from it a solution for which the
transmission coefficient has no poles in the upper complex half-plane
({\it i.e.} no bound states). Then, by a sequence of transformations of type
(\ref{zet1}-\ref{la3}) with suitably chosen parameters one can find a
solution for which the reflection to the left coefficient has no poles or
zeros in the upper half plane. 

We will call this solution of the phase reconstruction problem
{\it the left reference solution}. The left reference solution's
transmission coefficient, $T^{(r-)}(k)$, and the reflection to the left
coefficient, $R_-^{(r-)}(k)$, are analytic and have no zeros in $\Im(k)>0$.
In a similar way, we can define the right reference solution,
for which $T^{(r+)}(k)$ and $R_+^{(r+)}(k)$, are analytic and have no
zeros in $\Im(k)>0$.  The left and right reference solutions are connected
by a transformation of type (\ref{lp1}-\ref{lp3}):
\begin{eqnarray}
T^{(r+)}(k)&=& T^{(r-)}(k)=T^{(r)}(k), \label{ref1}\\
R_-^{(r+)}(k)&=&\frac{S_{N+M}(k)}{S_{N+M}(-k)}R_-^{(r-)}(k),\\
R_+^{(r+)}(k)&=&\frac{S_{N+M}(-k)}{S_{N+M}(k)}R_+^{(r-)}(k).
\label{ref3}\end{eqnarray}
Here, the polynomial
\begin{equation}
S_{N+M}(k)=A_N(k)B_M(-k),
\label{polAB}\end{equation}
where the zeros of the polynomial $A_N(k)\, /\, B_M(k)$ coincide (including
multiplicities) with the poles/zeros of $R_-^{(r-)}(k)$ in $\Im(k)>0$.
Taking into account(\ref{ro-r}) the polynomials are completely determined.

The reference solutions are in a certain sense the maximally non-symmetric
solutions. Indeed, as shown in the next section \ref{rat}, the potentials
corresponding to the left/right reference solutions are identically
zero for $x<0$ / $x>0$.                              
Other solutions of the phase reconstruction problem can be obtained from the
reference ones by adding bound states and reflection resonances with
reflectance-preserving transformations of type (\ref{lam1}-\ref{la3}). 

The left reference solution's $R_-^{(r-)}(k)$ is analytic and has
no zeros in $\Im(k)>0$. Thus, the logarithm  $\ln[R_-^{(r-)}(k)]$
is also analytic in $\Im(k)>0$.
The phase of the left reference solution (which is equal to the imaginary
part of $\ln[R_-^{(r-)}(k)]$) can be obtained from the logarithm of
its absolute value (which is equal to the real part of $\ln[R_-^{(r-)}(k)]$),
using a (subtracted) logarithmic dispersion relation (See {\it
e.g.} \cite{clinton,RL96}).

We will proceed in a different manner, which is more adequate with
the physical context.

The effective mass approximation is valid
only for energies within an interval not exceeding several hundred
millielectronvolts (meV) near the $\Gamma$ point minimum in the
$Al_xGa_{1-x}As$ system (or in the lattice matched $In_{1-x-y}Al_xGa_yAs$ 
systems). Thus, two sets of scattering data having the same (or close) low and intermediate-energy behavior, but whose exact high-energy behavior
is different, can be considered equivalent.
We need a good approximation of the SD in the physically relevant
range of energies, which obeys the high-energy constraints set forward
in the Appendix \ref{appA}. The Pad\'e approximation method\cite{baker} is
a good framework for that. An added bonus, which will be apparent
in the following section \ref{rat}, is the simplification of the calculations
needed for recovering the potential.

\subsection{Pad\'e phase reconstruction}\label{pha-3}

We start with approximating the input design data for the reflectance
by a type II [p,p+q+2] Pad\'e approximant\cite{baker}:
\begin{equation}
\calR(E)=\frac{\calP_{p}(E)}{\calQ_{p+q+2}(E)},
\label{R_E}\end{equation}
where $\calP_{p}(E)$ and $\calQ_{p+q+2}(E)$ are polynomials of 
degrees $p$ and, respectively, $p+q+2$ with $p,\,q\ge 0.$ 
Since the reflectance must be non-negative and less than 1 for all $E\ge 0$, 
the polynomials must satisfy
\begin{equation}
0\le\calP_{p}(E)<\calQ_{p+q+2}(E),
\label{R_E1}\end{equation}
on the positive half-axis. The second inequality (\ref{R_E1}) is strict 
for all $E>0$ and becomes an equality only for $E=0$. This ensures that 
$\calR(0)=1$ as it should be in the generic case\cite{calR0}.
We can rewrite it in the form 
\begin{equation}
\calQ_{p+q+2}(E)=\calP_{p}(E)+E\calK_{p+q+1}(E),
\label{R_E2}\end{equation}
where the polynomial $\calK_{p+q+1}(E)>0$ for $E\ge0$. 
The polynomial  $\calP_p(E)$ is normalized by setting 
\begin{equation}
\calP_{p}(0)=1.
\label{R_E3}\end{equation}
Finally, the transmittance is approximated by the [p+q+2,p+q+2] Pad\'e 
approximant:
\begin{equation}
\calT(E)=1-\calR(E)=\frac{E\calK_{p+q+1}(E)}{\calP_p(E)+E\calK_{p+q+1}(E)}.
\label{R_E4}\end{equation} 

We can choose the coefficients of the polynomials in (\ref{R_E})
as the parameters. These can be obtained from standard type II Pad\'e
fitting routines.  
Taking into account (\ref{R_E1}) with equality at $E=0$ and
(\ref{R_E3}), the fit is obtained by solving a system of $2p+q+2$
linear equations with $2p+q+2$ unknowns (the polynomials' coefficients),
which make the fit exact at $2p+q+2$ chosen points.
Thus, the  [p,p+q+2] Pad\'e (\ref{R_E}) is fully determined.
Eq. (\ref{R_E4}) does not introduce additional parameters. 

As we will see further on, it is advantageous to reparameterize
in terms of the zeros and poles of (\ref{R_E}). Since the reflectance and
the transmittance are real and non-negative, the zeros and poles of
(\ref{R_E}) and (\ref{R_E4}) are either real negative or come in
complex conjugate pairs.
The only exception to this rule are eventual real positive zeros of 
$\calP_p(E)$, which have an even order of degeneracy (generically =2).
The heterostructure is transparent to Bloch waves (maximal transmission
resonances) at the energies corresponding to these degenerate zeros.

Now, we want to find the left reference solution by solving eq. (\ref{refl}), 
\begin{equation} 
\left\vert R_-^{(r-)}(k)\right\vert^2=\calR(E), 
\label{refl-}\end{equation}
with the reflectance given by (\ref{R_E}). 
The reflection to the left coefficient, $R_-^{(r-)}(k)$, will be sought
as a [p,p+q+2] Pad\'e approximant in the variable $k$
\begin{equation}
R_-^{(r-)}(k)=-\frac{P_{p}(k)}{Q_{p+q+2}(k)},
\label{R-}\end{equation}
with the normalization 
\begin{equation}
P_{p}(0)=Q_{p+q+2}(0)=1, 
\end{equation}
which agrees with (\ref{R_E3}). As mentioned above, $R_-^{(r-)}(k)$ for
the left reference solution of (\ref{refl-}) is analytic in the upper
half-plane and has no zeros there.
Then, the zeros of both the denominator and numerator of (\ref{R-}) must
lie in the lower complex half-plane. A further constraint on the roots
follows from the relation (\ref{8}), 
\begin{equation}
\left[R_-(k)\right]^*=R_-(-k).\label{c-c}\end{equation}  
This relation holds only if for each $r$, which is a zero/pole of (\ref{R-}), 
$-r^*$ is also a zero/pole.

Let us now use (\ref{ro-r}) to represent all the polynomials
involved in the equation (\ref{refl-}) as products:
\begin{equation}
\frac{\prod_{i=1}^p(1-k/p_i)(1+k/p_i)}
{\prod_{j=1}^{p+q+2}(1-k/q_j)(1+k/q_j)}=
\frac{\prod_{i=1}^p(1-E/{\frak{p}}_i)}
{\prod_{j=1}^{p+q+2}(1-E/{\frak{q}}_j)}.
\end{equation}
Here $p_i$ and ${\frak{p}}_i, ~i=1,\ldots,p$ are the zeros of $P_p(k)$,
respectively $\calP_p(E)$, while $q_j$ and ${\frak{q}}_j, ~j=1,\ldots,p+q+2$
are the zeros of $Q_{p+q+2}(k)$, respectively $\calQ_{p+q+2}(E)$.

Taking into account the relation (\ref{ksq}), $E=\hbar^2k^2/m_e$, we find 
the following relations between the zeros of the polynomials:
\begin{eqnarray}
{\frak{p}}_i&=&\frac{\hbar^2p_i^2}{2m_e};~~i=1,\ldots,p;\label{p1}\\
{\frak{q}}_j&=&\frac{\hbar^2q_j^2}{2m_e};~~j=1,\ldots,p+q+2.\label{q1}
\end{eqnarray}
These relations solve the problem up to the ambiguity of the
signs of the square roots. Let us show that the reference solution
is unique. 

As mentioned in the Introduction, the square-root function as maps the
complex plane cut along $[0,+\infty)$ onto the upper complex half-plane,
{\it i.e.} if $E$ is in the cut plane, then $\Im\left(\sqrt{E}\right)>0$.
Since the zeros and poles of the reference solution cannot lie in the upper
half-plane the sign choices in (\ref{p1}-\ref{q1}) are
unique\cite{confluence}:
\begin{eqnarray}
\hbar p_i&=&-\sqrt{2m_e{\frak{p}}_i};~~i=1,\ldots,p;\label{p2}\\
\hbar q_j&=&-\sqrt{2m_e{\frak{q}}_j};~~j=1,\ldots,p+q+2.\label{q2}
\end{eqnarray}

It remains to show that if $r$ is one of the zeros, respectively poles, 
of (\ref{R-}), then $-r^*$ is also a zero, respectively pole. 
This is obvious if $r$ is real\cite{confluence} or imaginary. Otherwise,
both the real and the imaginary parts of $r^2$ are nonzero.
Then, $r^{*2}$ is also a zero (pole) of (\ref{R_E}), since as noted
above these come in complex-conjugate pairs. But our definition of
the square-root function leads to
$\sqrt{r^{*2}}=-\left(\sqrt{r^2}\right)^*.$

Thus, (\ref{R-}) with the zeros and poles given by (\ref{p2}-\ref{q2}) is 
the unique left reference solution for the reflection to the left coefficient
if the reflectance is given by the Pad\'e approximant (\ref{R_E}).
We can readily recover the complex transmission coefficient of the reference 
solutions using the same approach. 

The (complex) transmission coefficient of the reference solution is 
given by the [p+q+2,p+q+2] (diagonal) Pad\'e approximant
\begin{equation}
T^{(r)}(k)=\frac{kK_{p+q+1}(k)}{Q_{p+q+2}(k)},
\label{T-pade}\end{equation}
normalized by setting the coefficient of $k^{p+q+1}$ in the polynomial 
$K_{p+q+1}(k)$ equal to the coefficient of $k^{p+q+2}$ in the already 
determined polynomial $Q_{p+q+2}(k)$, so that $\lim_{k\to\infty}T(k)=1.$ 
Let 0 and ${\frak{k}}_j, ~j=1,\ldots,p+q+1$ be the zeros of the 
transmittance, (\ref{R_E4}).
\begin{equation}
\calP_p({\frak{k}})=\calQ_{p+q+2}({\frak{k}}).
\end{equation}
Repeating the reasoning that led to (\ref{p2}), we recover the zeros 
$\kappa_j$ of the polynomial $K_{p+q+1}(k)$
\begin{equation}
\hbar \kappa_j=-\sqrt{2m_e{\frak{k}}_j};~~j=1,\ldots,p+q+1,\label{kt}
\end{equation}
and
\begin{equation}
K_{p+q+1}(k)=-\frac{\prod_{i=1}^{p+q+1}\left(\kappa_i-k\right)}
{\prod_{j=1}^{p+q+2}q_j}.\label{Kp+q}
\end{equation}

Finally, using the relation (\ref{7ac}), we recover the left reference 
solution's $R_+^{(r-)}(k)$ as a [2p+q+1,2p+2q+3] Pad\'e approximant:
\begin{equation}
R_+^{(r-)}(k)=-\frac{K_{p+q+1}(k)}{K_{p+q+1}(-k)}
\frac{P_p(-k)}{Q_{p+q+2}(k)}.
\label{R+}\end{equation}
The first fraction in the above expression is a phase factor.

Thus, if the reflectance is given in the Pad\'e form (\ref{R_E}),
the left reference solution is uniquely determined. The right/left reflection
coefficients of the right reference solution are given by
right-hand side of (\ref{R-})/(\ref{R+}). The potentials corresponding to
the right/left reference solutions are mirror images:
$V^{(r-)}(x)=V^{(r+)}(-x)$.

Other solutions can be obtained by using transformations of type
(\ref{lam1} - \ref{lam3}) to introduce bound states into the reference
solutions.
Then, transformations of type (\ref{zet1} - \ref{lp3}) can be used for
reflection resonance management.
A distinguished class of solutions, having the same transmission
coefficient as the reference ones and involving no additional parameters, can
be obtained by redistributing reflection resonances between the reflection
coefficients to the left/right. This is achieved by factorizing the
polynomials  $P_p$ and $K_{p+q+1}$
\begin{equation}
P_{p}(k)=P_{p_+}(k)P_{p_-}(k);
~~~K_{p+q+1}(k)=K_{n_+}(k)K_{n_-}(k),
\label{C-k}\end{equation}
into factors which satisfy the complex conjugation relation (\ref{c-c})
for real $k$. Here, $p_++p_-=p$, $n_++n_-=p+q+1$ and all the factors
are normalized by $P_{p_\pm}(0)=1$.
Applying the transformation (\ref{lp1}-\ref{lp3}) with
$S(k)=K_{n_-}(k)P_{p_+}(-k)$ to the left reference solution, we obtain
a solution having the same transmission coefficient $T^{(r)}(k)$, given by
(\ref{T-pade}), and the reflection coefficients are given by the
[n$_\pm$+p,n$_\pm$+p+q+2] Pad\'e approximants
\begin{eqnarray}
R_-(k)&=&-\frac{K_{n_-}(k)}{K_{n_-}(-k)}
\frac{P_{p_+}(-k)P_{p_-}(k)}{Q_{p+q+2}(k)},\label{dis2}\\
R_+(k)&=&-\frac{K_{n_+}(k)}{K_{n_+}(-k)}
\frac{P_{p_+}(k)P_{p_-}(-k)}{Q_{p+q+2}(k)}.
\label{dis3}\end{eqnarray}

\section{Solution of the inverse scattering problem for the Schr\"odinger
equation  when the scattering data are rational functions}\label{rat}

In this section we will present the solution of the inverse scattering 
problem for the one-dimensional Schr\"odinger equation in the case when 
the scattering data are given in the  Pad\'e approximant form we
obtained in the preceding section \ref{pha-3}.
The inverse scattering problem in the case of rational coefficients has
been first considered by Kay and Moses\cite{K55,KM56a,KM56b,K60}. Significant 
results are due to Sabatier\cite{S83,CS}. We will follow
Sabatier's approach\cite{S83,CS} quite closely.

As shown in the Appendix \ref{i-M}, the potential in (\ref{2s}) can be
recovered from the transformation kernels.
\begin{equation}
V(x)=\mp2\frac{\rmd}{\rmd x} K_\pm(x;x\mp 0).\label{3rec}
\end{equation}
The transforming kernels $K_\pm(x;y)$ are the  solutions of the Marchenko
equations, (\ref{A14}), (\ref{A14plus})
\begin{eqnarray}
K_-(x;y)+M_-(x+y)&\hbox{\hskip -1pt}=\hbox{\hskip -1pt}&
\int_{-\infty}^x\hbox{\hskip -3pt}\rmd s M_-(y+s)K_-(x;s)=0,\nonumber\\
\label{m14-}\\
K_+(x;y)+M_+(x+y)&\hbox{\hskip -1pt}=\hbox{\hskip -1pt}&
\int^{\infty}_x\hbox{\hskip -3pt}\rmd s M_+(y+s)K_+(x;s)=0.\nonumber\\
\label{m14+}
\end{eqnarray}
The Marchenko kernels, $M_\pm(u)$, are given by (\ref{A15}).
\begin{equation}
M_\pm(u)={1\over{2\pi}}\int_{-\infty}^{+\infty}\rmd\kappa 
\rme^{\pm\rmi\kappa u} R_\pm(\kappa)
+\sum_{j}\left(C_j^\pm\right)^{-2}\rme^{\mp\lambda_ju}.\label{3m15}
\end{equation}
Here, $C_j^\pm$ are the normalization constants of the bound states, defined
in (\ref{b-state}).

Inspection of (\ref{m14-}) and (\ref{m14+}) shows that the first
variable, $x$, enters the equations only as a parameter. Also,
for negative/positive $x$, the integral in (\ref{m14-})/(\ref{m14+}) 
involves only negative/positive values of $s$. Since we need
only the contact values of the transformation kernels in (\ref{3rec}),
it is natural to use the $K_-$ version of (\ref{3rec}), obtained by
solving (\ref{m14-}) for negative $x$,  and the solution of (\ref{m14+})
for positive $x$. Taking into account the definition of the
Marchenko kernels, (\ref{3m15}), we see that
we can close the integration contour into $\Im(k)>0$ in the
expression of the relevant kernel ($M_+/M_-$ for $x>0/x<0$).
In particular, if one of the reflection coefficients is analytic there,
its contribution to the corresponding Marchenko kernel is identically zero.
In the absence of bound states, this implies that the corresponding
$M_\pm=0$ and, therefore, the potential will be zero on the corresponding
half-axis.

If the reflection coefficients are rational functions, like the
solutions of the phase reconstruction problem we obtained in
section \ref{pha-3}, they have at most a finite number of poles in $\Im(k)>0$.
The corresponding Marchenko kernels (\ref{3m15}) will be given by 
finite sums of exponentials, which go to zero at the corresponding infinity.
For simplicity's sake, we will assume that all the poles are simple.
The multiple pole case can be dealt with as a limiting case
of pole confluence. Then, the corresponding kernel, $M_+(u)/M_-(u)$ for
$u>0/u<0$, is
\begin{equation} 
M_\pm(u)=\rmi\sum_{j\in\Omega_\pm}\varrho_j^\pm
\rme^{\pm\rmi \nu_j^\pm u}.  \label{2-18}
\end{equation}
Here the sets of (all distinct) complex numbers
$\left\{\nu_j^\pm\right\}_{j\in\Omega_\pm}$ consist of the (simple) poles of
$R_\pm(k)$ in $\Im(k)>0$ and, if bound states with energies
$-\hbar^2\lambda_j^2/2m_e$ are present,  $\rmi\lambda_j$.
The coefficients $\varrho_j^\pm$ are either the residues of
the corresponding reflection coefficients or, if $\nu_j^\pm$ comes
from a bound state, $-\rmi\left(C_j^\pm\right)^{-2}$.

Substituting the separable Marchenko kernels (\ref{2-18}) into
the Marchenko equation (\ref{m14-}) for $x<0$, we obtain
\begin{equation}
K_-(x;y)=\rmi\sum_{j\in\Omega_-}\rho^-_j\rme^{-\rmi\nu_j^-y}
%\left[\rme^{-\rmi\nu_j^-x}+
Y_j^-(x),%\right],
\label{2-19}\end{equation}
where 
\begin{equation}
Y_j^-(x)=\rme^{-\rmi\nu_j^-x}+ 
\int_{-\infty}^x\rmd y K_-(x;y)\rme^{-\rmi\nu_j^-y}.\label{2-Yj}
\end{equation}
Multiplying (\ref{2-19}) by $\rme^{-\rmi\nu_m^-x}$ and integrating with
respect to $y$ we obtain a system of $\#(\Omega_-)$ linear equations for
the $\#(\Omega_-)$ unknowns $Y_j^-(x)$. Here we used the notation
$\#(\Omega)$ for the number of elements in the set $\Omega$. Substituting
the solution into (\ref{2-19}), we obtain after a little algebra
\begin{equation}
K_-(x;x-0)=-\frac{\rmd}{\rmd x} \Tr
\ln \left[1-\rme^{-\rmi x\sfN^-}\sfD^-\rme^{-\rmi x\sfN^-}\right]. %\ln \det \sfD^-(x) =
\label{2-20}\end{equation}
Here, $\sfN^-$ and $\sfD^-$ are $\#(\Omega_-)$ square matrices with
\begin{eqnarray}
N_{mj}^-&=&\delta_{mj}\nu_j^-;\\
D^-_{mj}&=&\frac{\rho_j^-}{\nu_m^-+\nu_j^-}.
\label{2-sys1}\end{eqnarray}

Solving in the same manner the equation for $K_+(x;y)$, we obtain
\begin{equation}
K_+(x;x+0)=\frac{\rmd}{\rmd x}\Tr %\ln [1-\sfQ^+(x)\sfD^+\sfQ^+(x)],
\ln \left[1-\rme^{-\rmi x\sfN^+}\sfD^+\rme^{-\rmi x\sfN^+}\right]. 
\label{2-22}\end{equation}
where the elements of the $\#(\Omega_+)$ square matrices  $\sfN^+$
and $\sfD^+(x)$ are
\begin{eqnarray}
N_{mj}^+&=&\delta_{mj}\nu_j^+;\\
D^+_{mj}&=&\frac{\rho_j^+}{\nu_m^++\nu_j^+}.
\label{2-sys1+}\end{eqnarray}

We can substitute now (\ref{2-20}) and (\ref{2-22}) into (\ref{3rec})
to obtain
\begin{equation}
V(x)=-8\Tr \left[\sfN^\pm\frac 1{1-\sfQ^\pm(x)}
\sfN^\pm\frac{\sfQ^\pm(x)}
{1-\sfQ^\pm(x)}\right],
\label{2-23}\end{equation}
where the $\pm$ signs are for $x>0$/$x<0$ and 
\begin{equation}
\sfQ^\pm(x)=\rme^{2\rmi|x|\sfN^\pm}\sfD^\pm.
\label{2-24}\end{equation}

Thus, the potential is given by (\ref{2-23}) in terms of a trace of
the inverse of finite matrices.

\section{Variable-mass mapping: Schr\"odinger's to BenDaniel and
Duke's equation}\label{vmm}

In  section \ref{rat} we have presented a simple algorithm
for solving the inverse scattering problem for the one-dimensional
Schr\"odinger equation in the case of rational SD. %%, which are obtained %in our approach to the phase reconstruction problem.

If the conduction electron dynamics is described by the BDD equation, we 
will not start by posing the inverse scattering problem for that equation 
from scratch.
Instead of that, we will use a family of unitary 
transformations\cite{BBM95,BBM96,BMMV97}. The transformations 
map the one-dimensional Schr\"odinger equation,
(\ref{0-s1}) with $\mbq_\perp=0$, and potential $V_S(z)$,
\begin{equation}
\psi^{\prime\prime}(z)
+\left[k^2-V_S(z)\right]\psi(z) =0,
\label{3-1} \end{equation}
into BenDaniel and Duke equations, (\ref{0-bdd1}), with $\mbq_\perp=0$,
and potential $V_{BDD}(z)$,
\begin{equation}
\left[\frac{m_\infty}{m(z)}\psi^{\prime}(z)\right]^\prime
+\left[k^2-V_{BDD}(z)\right]\psi(z) =0,
\label{3-2}\end{equation}
with a variable effective mass $m(z)$ and a potential $V_{BDD}(z)$.
The effective mass and the new potential are functionally related
to the parameters defining the unitary transformation. 
Since the mapping is unitary, the SD for the equation (\ref{3-2})
will be identical with the SD for (\ref{3-1}). 
We start thus with a Schr\"odinger reference equation, (\ref{3-1}),
which is the solution of the inverse scattering problem. From it
we obtain a family of BDD equations with the same scattering  
data. The problem is to choose among these transformations those 
which map the solution of the inverse problem for the SE onto 
acceptable BDD equations.

\subsection{Unitary mapping}\label{umaping}
Let us introduce a (nonlinear) coordinate transformation
\begin{equation}
z=X(x),
\label{3-3}\end{equation}
which maps the interval $(-\infty,+\infty)$ into $(-\infty,+\infty)$.
Here the function $X(x)$ is a smooth monotonically increasing function: 
$X^\prime(x)>0$.
The monotonicity ensures that a unique inverse transformation exists:
$x=Z(z)$, with 
\begin{equation}Z[X(x)]=x\end{equation} 
and $X[Z(z)]=z$. The inverse function $Z(z)$ is also smooth and 
monotonically increasing, with 
\begin{equation}
Z^\prime(z)=1/X^\prime[Z(z)].
\label{3-4}\end{equation}

Let us associate with the coordinate transformation (\ref{3-3}) a mapping
$\hat{\sf{U}}_X$ of the space of square-integrable functions, $L^2$, which transforms 
each element of the space $f\in L^2$ into $\hat{\sf{U}}_Xf$ with 
\begin{equation}
(\hat{\sf{U}}_Xf)(x)=\sqrt{X^\prime(x)}f[X(x)].\label{3-5}
\end{equation} 
Making the change of variables (\ref{3-3}) in the normalization integral
\begin{eqnarray}
\langle f|f\rangle&=&\int_{-\infty}^{+\infty}\rmd z |f(z)|^2=
\int_{-\infty}^{+\infty}\rmd x X^\prime(x)|f[X(x)]|^2\nonumber\\
&=&\langle \hat{\sf{U}}_Xf|\hat{\sf{U}}_Xf\rangle
=\langle f|\hat{\sf{U}}_X^*\hat{\sf{U}}_X|f\rangle,\label{3-6}
\end{eqnarray} 
we see that the mapping $\hat{\sf{U}}_X$ maps the square integrable functions 
into square integrable functions conserving the norm, {\it i.e.} 
$\hat{\sf{U}}_X$ is isometric on $L^2$. Since the inverse transformation,
\begin{equation}
\left(\hat{\sf{U}}_X^{-1}f\right)(z)=\sqrt{Z^\prime(z)}f[Z(z)].\label{3-7}
\end{equation} 
exists and is non-singular, the mapping $\hat{\sf{U}}_X$ is unitary: 
\begin{equation}
\hat{\sf{U}}_X^*=\hat{\sf{U}}_X^{-1}.\label{3-U}
\end{equation} 

We want to see the effect of the transformation $\hat{\sf{U}}_X$ on functions
satisfying the Schr\"odinger equation (\ref{3-1}). We will consider
only the coordinate transformations for which the function $X(x)$
is twice differentiable with a piecewise continuous third
derivative\cite{smoo3}, $X^{\prime\prime\prime}$.
Substituting (\ref{3-5}) into (\ref{3-1}), we obtain after some algebra, 
\begin{equation}
\left[\frac{\chi^\prime(x)}{[X^\prime(x)]^2}\right]^\prime+
\left[k^2-W_X(x)\right]\chi(x)=0.
\label{3-8}\end{equation}       
Here $\chi(x)=(\hat{\sf{U}}_X\psi)(x)$ and
\begin{equation}
W_X(x)=V_S[X(x)]+
\frac{X^{\prime\prime\prime}(x)}{2[X^\prime(x)]^3}-
\frac{5[X^{\prime\prime}(x)]^2}{4[X^\prime(x)]^4}.
\label{3-9}\end{equation}

The equation (\ref{3-8}) satisfied by the transformed function 
$\chi(x)$ resembles the BDD equation (\ref{3-2}) if we set
\begin{equation}
m(x)=m_\infty\left[X^\prime(x)\right]^2=m_{cond}[c(x)].
\label{3-10}\end{equation}
The function $X(x)$ must satisfy 
\begin{equation}
\lim_{x\to\pm\infty}X^\prime(x)=1.
\label{3-11}\end{equation}
This ensures that $m(x)$ tends at infinity to the constant limit $m_\infty$.
We will also require that $X^\prime(x)-1$ decays at infinity faster
than $|x|^{-2-\delta}$ for some $\delta>0$. In this case, 
the limits of $X(x)-x$ at $\pm\infty$ are finite. This 
ensures that the Jost solutions of the original Schr\"odinger equation
are mapped into Jost-type solutions of the transformed one. 

Let $f_\pm(x;k)$ be the Jost solutions of (\ref{3-1}),
which obey the boundary conditions (\ref{A3}):
\begin{equation}
\lim_{x\to\pm\infty}\rme^{\mp\rmi kx}f_\pm(x;k)=1.
\label{3-jbc}\end{equation}
Then, applying to them the mapping (\ref{3-5}), we obtain
near the corresponding infinities
\begin{eqnarray}
\lim_{x\to\pm\infty}\rme^{\mp\rmi kx}\bigl({\sf{\hat{U}}}_X f_\pm\bigr)(x;k)=
\rme^{\rmi kd_\pm}.
\label{s-s}\end{eqnarray}
Here, 
\begin{equation}%{eqnarray}%
d_+=\int_0^{+\infty}\rmd \xi \left[X^\prime(\xi)-1\right];~~
d_-=\int^0_{-\infty}\rmd \xi \left[X^\prime(\xi)-1\right];
\label{3-dpm}\end{equation}%{eqnarray}%
where we assumed $X(0)=0$.
Let $f_\pm^{(X)}(x;k)$ be the Jost-type solutions of the transformed equation
(\ref{3-8}), defined by the same boundary conditions (\ref{3-jbc}).
Taking into account the asymptotic behaviors of the Jost functions
of the Schr\"odinger equation (\ref{3-1}), we find the relation between the 
scattering data for the transformed equation and those of the original one:
\begin{eqnarray}%{equation}
T^{(X)}(k)&=&\rme^{\rmi k(d_+-d_-)}T(k),\\
R^{(X)}_\pm(k)&=&\rme^{\pm 2\rmi kd_\pm}R_\pm(k).
\label{3-newj}\end{eqnarray}%{equation}

Thus, the reflectances of the original and the
transformed equations are equal.

In the following subsection we will study the variable effective
mass mapping nonlinear differential equation in the general case
of an arbitrary dependence of the band offset on the concentration.
We will show that the physically realizable concentration profiles
for devices embedded in a material of homogeneous composition 
can be obtained by solving a (nonlinear) boundary value problem.
A necessary condition for the existence of achievable solutions is that 
the dependence of the conduction band offset ${\cal{E}}_c$ on the
effective mass be a {\it non-decreasing} one. If this condition
is not satisfied, it may be still possible to embed the device in
a {\it periodic} super lattice, a case which will not be discussed
in this paper. In section \ref{bv-sol} we consider in more detail
the case when ${\cal{E}}_c(m)$ is a linear function, an approximation
which seems quite reasonable in the $AlGaAs$ system. In this case, the
solution of the nonlinear boundary value problem can be then expressed
through the canonical solutions of the Cauchy (initial value)
problem for a third-order linear differential equation. We also discuss
methods for obtaining stable approximate solutions.

\subsection{Differential equation}\label{diff}

Let us now compare the potentials $W_X$, (\ref{3-9}), and $V_{BDD}$.
The latter is given by (\ref{S2}) and (\ref{ksq}):
\begin{equation}
V_{BDD}(x)=\frac{2m_\infty}{\hbar^2}\left[
\calE(x)-\calE(\infty)+\Phi_{sc}(x)\right].
\label{3-12}\end{equation}
Here, the band offset is %a function of the local chemical composition
\begin{equation}
\calE(x)=\calE_{cond}[c(x)],
\label{3-13}\end{equation}
and we set $U_{ext}(x)=0$. The zero of the energy scale is chosen at
the value of the band offset for the embedding (asymptotic) composition.

The band offset, $\calE(x)$ and the effective mass, $m(x)$
depend on the position only through their dependence on the local chemical
composition $c(x)$, (\ref{3-10}) and (\ref{3-13}). 
We will consider only the case when the function $m(c)$ is invertible. 
Then, we can substitute the inverse function $c(m)$ into $\calE_{cond}(c)$:
\begin{equation}
\calE_{cond}(c)=\calE_c(m).
\label{3-14}\end{equation}

In this section we consider only unbiased structures with negligible
electron density.
Then, in the square brackets in (\ref{3-12}), $\Phi_{sc}(x)$ is
equal to zero. Equating the two potentials (\ref{3-9}) and (\ref{3-12})
an using (\ref{3-10}), we obtain a third-order nonlinear differential
equation for $X(x)$:
\begin{eqnarray}
\frac{X^{\prime\prime\prime}(x)}{2[X^\prime(x)]^3}&-&
\frac{5[X^{\prime\prime}(x)]^2}{4[X^\prime(x)]^4}
+V_S[X(x)]=\label{3-15}\\
&=&\frac{2m_\infty}{\hbar^2}\left\{
\calE_c[m_\infty X^{\prime\, 2}(x)]-\calE_c(m_\infty)\right\}.\nonumber
\end{eqnarray}
The values of $X^\prime(x)$ are restricted to the physically achievable
interval
\begin{equation}
\sqrt{\frac{\underline{m}}{m_\infty}}\le\dot{X}
\le\sqrt{\frac{\overline{m}}{m_\infty}}.
\label{3-16}\end{equation} 
Here $\underline{m}$ and $\overline{m}$ are the minimal and, respectively,
the maximal values of the effective mass, $m(c)$, in the physically achievable
chemical composition range.

The equation (\ref{3-15}) does not depend explicitly on  $x$. Setting
\begin{equation}
X^\prime=S\left[X(x)\right],
\label{3-17}\end{equation}
substituting this into (\ref{3-15}) and replacing the derivatives 
with respect to $x$ according to
\begin{equation}
\frac{\rmd}{\rmd x}=\frac{\rmd X}{\rmd x}
\frac{\rmd}{\rmd X}=S\frac{\rmd}{\rmd X},
\label{3-18}\end{equation}
we obtain a second-order equation for $S,$ 
\begin{equation}%\begin{eqnarray}
2\frac{S^{\prime\prime}}{S}-3\frac{\left (S^{\prime}\right )^2}{S^2}
=\frac{8m_\infty}{\hbar^2}\left[\calE_c(m_\infty S^2)-
\calE_c(m_\infty)-U(X)\right].
\label{3-19}\end{equation}%\end{eqnarray}
Here $S^\prime$ is the derivative of $S$ with respect to $X$,
$U(X)=\hbar^2V_S(X)/2m_\infty$ is the potential in energy units.
The asymptotic condition (\ref{3-11}) reads now $S(\pm\infty)=1$.

\subsection{Boundary value problem}\label{bv}
%\subsection{Behavior at infinity and achievability}\label{inf}

Let us consider potentials $V(X)$ which are identically equal to zero
outside some interval [$X_-,X_+$].
For $X$ outside this interval,
the equation (\ref{3-19}) does not depend explicitly
on $X$ so that its order may be further reduced. Setting 
\begin{equation}
m(X)=m_{\infty} S^2 \label{3-20}
\end{equation} 
and
\begin{equation}
S^\prime(X)=Q(m(X)),\label{3-21}
\end{equation}
the equation (\ref{3-19}) becomes
\begin{equation}
2\frac{{\rmd} Q^2}{{\rmd} m} -3\frac{Q^2}{m}=
\frac{10}{\hbar^2}\biggl[\calE_c(m)-\calE_c(m_\infty)\biggr]. 
\label{3-22}\end{equation}
The solution of (\ref{3-22}) is 
\begin{equation}
Q^2(m)=\frac{4m^{\frac 32}}{\hbar^2}\int_{m_\infty}^m
\bigl[\calE_c(\mu )-\calE_c(m_\infty)\bigr]
\mu^{-\frac 32}\rmd\mu. \label{3-23}
\end{equation}
If $m$ is close to $m_\infty,$ the right hand side of
(\ref{3-23}) goes to zero as $(m-m_\infty)^2.$ 
Indeed, for small $|\mu -m_\infty|,$ the term in square 
brackets in (\ref{3-23}) is approximately equal to 
$\calE_c^{\prime}(m_\infty)\left (\mu -m_\infty \right)$ 
and
\begin{equation}
Q^2(m)\approx \frac{2\calE^\prime _c(m_\infty)}{\hbar^2}
\left (m-m_\infty\right )^2 +{\cal{O}}\left[(m-m_\infty)^3\right]. 
\label{3-24}\end{equation}

Since $Q^2$ is non-negative, inspection of (\ref{3-24}) shows 
that the effective mass may tend to a constant limit at 
infinity {\em if and only if} the band offset dependence 
on the effective mass, $\calE_c(m),$ is a 
{\em non-decreasing function}. 
\begin{equation}
\frac{\rmd\calE_c(m)}{\rmd m}\ge 0
\label{3-25}\end{equation}     
In plain words, if the effective mass does not follow the band
offset, then the potential $V(X)$ cannot be embedded in an alloy of 
homogeneous composition.  Noting that in this
case the device may be embedded in a periodic superlattice, 
we will restrict ourselves here to the case (\ref{3-25}). 

Let us note that equation (\ref{3-19}) may be solved in quadratures 
on the intervals $(-\infty,X_-)$ and $(X_+,+\infty)$.
Indeed, substituting (\ref{3-23}) into (\ref{3-21}) 
and taking into account (\ref{3-20}) we get 
\begin{equation}
\frac{\rmd S}{\rmd X}=\pm\hbox{sign}(1-S) Q(m_{\infty} S^2),
\label{3-26}\end{equation}
where $Q(m)$ is the non-negative square root of the 
right hand side of (\ref{3-23}) and the sign is positive on
$(X_+,+\infty)$ and negative on $(-\infty,X_-)$.

The solution on $(X_+,+\infty)$ is given in parametric form by
\begin{eqnarray}
X &=& X_++\hbox{\rm sign}\left[1-S(X_+)\right]
\int_{S_+(X_+)}^{s} \frac{\rmd \sigma}{Q(m_{\infty}\sigma^2)},\label{3-27}\\
x &=& x_++\hbox{\rm sign}\left[1-S(X_+)\right]
\int_{S_+(X_+)}^{s} \frac{\rmd \sigma}{\sigma Q(m_{\infty} \sigma^2)},
\label{3-28}\end{eqnarray}
where $x_+$ is the value of $x$ which maps to $X_+$: 
$X_+=X(x_+).$ A similar representation is valid for 
the interval $(-\infty ,X_-).$

Thus, we have obtained 
the solutions which are regular at infinity. There, 
$S(X)=1 +{\cal{O}}\left[\rme^{-\kappa|X|}\right]$ tends 
exponentially to the constant limit $1.$
The asymptotic rate of decay $\kappa$ may be obtained 
from (\ref{3-24}):
\begin{equation}
\kappa =
\frac{2m_\infty}{\hbar}\sqrt{2\calE_c^\prime(m_\infty)}.  
\label{3-29}\end{equation}

Let us now briefly discuss the asymptotic behavior at 
infinity for the regular solutions of (\ref{3-19}) for 
potentials $V(X)$ which go to zero at infinity.
One may readily see from (\ref{3-19}) that 
$S(X)$ must still converge to $1$ at infinity.  
As long as the potential $V(X)$ decays at infinity 
faster than $\rme^{-\kappa|X|}$ the properties of the solutions
discussed above remain asymptotically valid.
For potentials with slower falloff at infinity, the regular solution
tends asymptotically to the regular solution of the equation 
$S^{\prime\prime}=8m_{\infty}\calE_c^\prime(m_\infty)S/\hbar^2+V(X)/2.$

% \subsection{Boundary value problem}\label{bv}

It is important to note that any regular solution of (\ref{3-19})
satisfies at all points $X_\pm$ outside the support of the potential
the boundary conditions (\ref{3-26}). 
To construct a solution of (\ref{3-19}) which is regular 
on the whole axis $X,$ we will chose a value for the 
embedding alloy effective mass, $m_\infty$. 
The regular solutions on the interval $(-\infty ,X_-)$ can be parameterized
by the value $S(X_-)=\tau$. The value of the 
derivative at $X_-$ is given by (\ref{3-26}) with 
the minus sign: 
\begin{equation}
S^\prime(X_-;\tau)=
-\hbox{\rm sign}\left[1-\tau\right] Q(m_{\infty} \tau^2),  
\label{3-30}\end{equation}
where $Q(m)$ is the positive square root of the right-hand
side of (\ref{3-23}).

Then, we integrate the equation (\ref{3-19}) numerically with 
the initial values defined above over the support of the 
potential up to $X_+,$ obtaining  $S(X_+;\tau)$ and 
$S^\prime(X_+;\tau).$ 
If the solution is regular then it must satisfy (\ref{3-26}) 
with the plus sign  at $X_+$: 
\begin{equation}
S^\prime(X_+;\tau)=
\hbox{\rm sign}\left[1-\tau\right] 
Q\left[m_{\infty} S^2(X_+;\tau)\right]. 
\label{3-31}\end{equation}
If they exist, the solutions of equation (\ref{3-31}) give the values
of $\tau$ for which we can find regular solution of the differential
equation (\ref{3-19}) on the whole real axis. To be achievable these solutions
must also satisfy the physical bounds (\ref{3-16}).

Numerically, the shooting method outlined here is  rather
ill-conditioned. Since the general solution of (\ref{3-19}) is
singular, multiple-precision arithmetic has to be used
for the integration of the differential equation if the support of
the potential is not short enough.
In the following section we will examine in more detail
the important case when the dependence of the 
band offset on the effective mass is linear. We will show that
in this case the solutions of the nonlinear equation (\ref{3-19})
can be found among the solutions of a third-order {\em linear}
equation.

\subsection{Solution of the boundary value problem in the case of linear
dependence of the band offset on the effective mass}\label{bv-sol}

%\subsection{Linear dependence of the band edge offset on the effective mass}\label{lin}

For some alloys like $Al_cGa_{1-c}As$ in the concentration range
$0\le c\le .45$ the conduction band minimum is at the center of the 
Brillouin zone ($\Gamma$) and the dependence of $m_{cond}(c)$ and 
$\calE_{cond}(c)$ on the concentration $c$ approximately  
linear. In the $Al_cGa_{1-c}As$ system, for $0\le c\le .45,$  the offset
from the position at $GaAs$ is
\begin{equation}
\calE_c(m)={B}(m-\underline{m}).
\label{3-32}\end{equation}
The constant ${B}$ is
\begin{equation}
{B}= \frac{\Delta E}{\overline{m}-\underline{m}}=9.41 eV/m_0,
\label{3-33}\end{equation}
Here $\underline{m}=.067m_0 $ is the conduction band effective mass for $GaAs$;
$\overline{m}=.104m_0 $ and $\Delta E$ are, respectively, the mass and 
band offset for $Al_{.45}Ga_{.55}As$.  $m_{0} $ is the electron mass.

Substituting this into (\ref{3-19}) we obtain
\begin{equation}
2{S^{\prime\prime}}{S}-3{\left (S^{\prime}\right )^2}
=\kappa_\infty^2{S^2}\left[S^2-1-v(X)\right].
\label{3-34}\end{equation}
Here,
\begin{eqnarray}%{equation}
\kappa_\infty^2&=&{8{B}m_\infty^2}/{\hbar^2},\label{3-35}\\%~~
v(X)&=&{4V(X)}/{\kappa_\infty^2}=U(X)/({B}m_\infty),\label{3-36}
\end{eqnarray}%{equation}
and $U(X)=\hbar^2V(X)/2m_\infty$ is the potential measured in energy units.
%\subsection{Linear equation}

Let us define a new unknown function %$T(X)$ --- the inverse of $S(X)$:
\begin{equation}
T(X)= {3-3}/{S(X)}. 
\label{3-39}\end{equation}
Substituting (\ref{3-39}) into (\ref{3-34}) we obtain the equation
satisfied by $T(X)$:
\begin{equation}
2TT^{\prime\prime}-T^{\prime 2}+
\kappa_\infty^2\left\{1-\bigl[1+v\bigr]T^2\right\}=0.
\label{3-40}\end{equation}
Here and whenever it does not lead to ambiguities we will omit 
the arguments of the functions. Taking the derivative of (\ref{3-40})
we obtain a third-order linear equation for $T$:
\begin{equation}
T^{\prime\prime\prime}-\kappa_\infty^2\left(1+v\right)T^{\prime}
-\frac 12\kappa_\infty^2v^\prime T=0.
\label{3-41}\end{equation}
An arbitrary solution of (\ref{3-41}) will satisfy the second-order
nonlinear equation (\ref{3-40}) with the zero in the right hand side
replaced by some constant ${\calK}$:
\begin{equation}
2TT^{\prime\prime}-T^{\prime 2}-
\kappa_\infty^2\bigl[1+v\bigr]T^2={\cal K}.
\label{3-42}\end{equation}
If $\calK=-\kappa_\infty^2$, the solution satisfies also (\ref{3-40}).

Let $F_{\alpha\beta\gamma}(X;X_-)$ be the solution of the initial value
(Cauchy) problem for the linear equation (\ref{3-41}) satisfying the
initial conditions
\begin{eqnarray}
F_{\alpha\beta\gamma}(X_-;X_-)&=&\alpha,\nonumber\\
F_{\alpha\beta\gamma}^\prime(X_-;X_-)&=&\beta,\label{3-43}\\
F_{\alpha\beta\gamma}^{\prime\prime}(X_-;X_-)&=&\gamma,\nonumber
\end{eqnarray}
at the point $X_-$, which is a point of continuity of the potential $v(X)$.
Since  (\ref{3-41}) is linear,
\begin{equation}
F_{\alpha\beta\gamma}=\alpha F_{100}+\beta F_{010}+\gamma F_{001}.
\label{3-44}\end{equation}
The function $F_{\alpha\beta\gamma}(X;X_-)$ also satisfies the second-order
equation (\ref{3-42}) with
\begin{equation}
{\cal K}_{\alpha\beta\gamma}=2\alpha\gamma -
\beta^2 -\kappa_\infty^2\alpha^2\left[1+v(X_-)\right],
\label{3-45}\end{equation}
in the right hand side.

%\subsection{Solution of the boundary value problem in the case of linear dependence of the band offset on the effective mass}\label{bv-sol}

In \ref{bv} we outlined the shooting method for solving the boundary
value problem which leads to the regular (acceptable) solutions of
(\ref{3-19}). It is well known that shooting methods
are prone to numerical instabilities even for linear boundary
value problems. Another unpleasant feature is the fact that we have to
integrate the equation numerically over the support of the potential
for each value of $\tau$.

Let us state the boundary value problem for the function $T(X)$,
assuming the potential $v(X)$ to be identically zero outside the
interval $(X_-,X_+)$. The solutions regular on $(-\infty,X_-)$
and $(X_+,+\infty)$ are 
\begin{eqnarray}
T_-(X)&=&1+\left[T(X_-)-1\right]\rme^{\kappa_\infty(X-X_-)},\label{3-46}\\
T_+(X)&=&1+\left[T(X_+)-1\right]\rme^{-\kappa_\infty(X-X_+)}.
\label{3-47}\end{eqnarray}
Thus, the boundary conditions at $X_\pm$ which must be satisfied by
the regular solution are
\begin{eqnarray}
T^\prime(X_-)&=&\kappa_\infty\left[T(X_-)-1\right],\label{3-48}\\
T^\prime(X_+)&=&-\kappa_\infty\left[T(X_+)-1\right].
\label{3-49}\end{eqnarray}
Thus, we can parameterize the regular solutions (\ref{3-46}) by the value
of $T(X_-)=\alpha$. Then, $T^\prime(X_-)$ is given by (\ref{3-48}).
Then we integrate (\ref{3-40}) up to $X_+$, where $T$ and its derivative
must satisfy (\ref{3-49}), whence the acceptable values of
$\alpha$ are determined.

Now, we can use the linear equation satisfied by $T$  to express the
solution of the initial value problem through the canonical solutions
$F_{100},~ F_{010}$ and $F_{001}$ of the initial value problem
for (\ref{3-41}). Using (\ref{3-40}) with $v(X_-)=0$ to find
$T^{\prime\prime}(X_-)$, we find the solution satisfying (\ref{3-48})
\begin{eqnarray}
T_\alpha (X)&=&\alpha F_{100}(X;X_-)+\kappa_\infty (\alpha-1) F_{010}(X;X_-)
\nonumber\\&+&\kappa^2_\infty (\alpha-1) F_{001}(X;X_-).
\label{3-50}\end{eqnarray}
At $X_+$, $T_\alpha(X)$ must satisfy the boundary condition (\ref{3-49}).
Whence we find $\alpha$:
\begin{equation}
\alpha=1-\frac{F^\prime_{100}(X_+;X_-)
+\kappa_\infty\left[F_{100}(X_+;X_-)-1\right]}
{F^\prime_{1\kappa_\infty\kappa_\infty^2}(X_+;X_-)+
\kappa_\infty F_{1\kappa_\infty\kappa_\infty^2}(X_+;X_-)},
\label{3-51}\end{equation}
where $F_{\alpha\beta\gamma}$ was defined in (\ref{3-43}).
Thus, the solution of the boundary value problem for
the nonlinear equation (\ref{3-40}) is expressed through the
canonical solutions of the initial value problem at $X_-$ for the linear
equation (\ref{3-41}).

The value of $X_-$ can be chosen arbitrarily as long as the potential
$v(X_-)=0$. We may safely assume that $v(X)=0$ in a neighborhood of $X_-.$
The functions $F_{100}$ and $F_{1\kappa_\infty\kappa_\infty^2}$
are also solutions of
the second-order nonlinear equation (\ref{3-42}) with
${\cal K}$ determined from (\ref{3-45}).
We have found the value of $\alpha$, which determines 
the solution of the boundary value problem, through the solutions of two
second order (albeit nonlinear) equations. Although the integration
must be performed only once from $X_-$ to $X_+$, it is still numerically
unstable.

\subsection{Approximate solutions}\label{map-ap}

As mentioned above the numerical integration of the differential
equation over typical device lengths 20-40$nm,$ having the order of
magnitude of the electron mean free path, is rather ill conditioned.
Taking the values of the parameters for the $AlGaAs$ system, above (\ref{3-33}),
and a value $m_\infty\approx .1m_{\infty} $ for the effective mass of the
embedding material, we obtain that the natural length scale for the
differential equation (\ref{3-34}) is $1/\kappa_\infty\approx .3nm,$
which is comparable to the lattice period.

The equation (\ref{3-34}) can be rewritten as
\begin{equation}
S^2-1-v(X)=\frac{2{S^{\prime\prime}}{S}-3{\left (S^{\prime}\right )^2} }
{\kappa_\infty^2S^2}.
\label{3-60}\end{equation}
Since the right-hand side of (\ref{3-60}) has the small factor
$\kappa_\infty^{-2}$, one is tempted to proceed in a
a {\it na\"\i ve} "quasi-classical" way and neglect the right-hand side
entirely. Then,
\begin{equation}
S(X)\approx \sqrt{1+v(X)}, 
\label{3-61}\end{equation}
and finding the coordinate transformation $X(x)$ reduces to a simple
quadrature. This gives surprisingly reasonable results.

Finding  corrections to (\ref{3-61}) seems a rather tedious task,
especially for potentials which have discontinuities.
We will note instead that from (\ref{3-36}),
\begin{equation}
v(X)=\frac{U(X)}{{B}m_{\infty}} \approx
\frac{U(X)}{.94eV},\label{3-B}
\end{equation}
for the $AlGaAs$ system. Since typical potential values are $\pm$100-200$meV$,
the potential $v(X)$ (measured in the natural units of the problem)
is small compared to 1. A "small potential" perturbative approach to
solving the boundary value problem is thus indicated.

Let us assume that $v\sim \alpha $ and  seek $S$ as a series
\begin{equation}
S(X)=S_{(0)}(X)+\alpha S_{(1)}(X)+\alpha^2 S_{(2)}(X)+\ldots,
\label{3-62}\end{equation}
substitute $v\to\alpha v$ and (\ref{3-34}), rewritten as 
\begin{equation}
S^{\prime\prime}=3S^{\prime\,\, 2}/2S\, +\,
\kappa_\infty^2S\left[S^2 -1-v\right]/2\, ,
\label{3-63}\end{equation}
and expand into a power series in $\alpha$. Then, after equating the terms with 
the same power of $\alpha$ and setting $\alpha=1$, we obtain
a hierarchy of linear differential equations for the functions 
$S_{(\ell)}(X),$ $\ell=0,1,\ldots$.  

The first equation from  (\ref{3-62}) is  $S_{(0)}^{\prime\prime}(X)=0$.
The solution must go to $1$ when $X$ goes to infinity so that
$S_{(0)}(X)\to 1$, while all the other $S_{(\ell)}(X)\to 0$ as
$X\to\pm\infty.$  Then,         
\begin{equation}
S_{(0)}(X)=1.
\label{3-64}\end{equation}
For $\ell\ge 1$, the hierarchy has the form
\begin{eqnarray}
S_{(1)}^{\prime\prime}&-&\kappa_\infty^2S_{(1)}=
-\frac 12 \kappa_\infty^2v;\label{3-65}\\
S_{(2)}^{\prime\prime}&-&\kappa_\infty^2S_{(2)}=
\frac 32 S_{(1)}^{\prime\, 2}%\nonumber\\&+&
+\frac 12\kappa_\infty^2S_{(1)}\left[3S_{(1)}-v\right];\label{3-66}\\
&\ldots&\nonumber\\
S_{(\ell)}^{\prime\prime}&-&\kappa_\infty^2S_{(\ell)}= %\nonumber\\&\phantom{-}&
{\cal{F}}_\ell\left(X;S_{(1)},\ldots ,S_{(\ell-1)},v\right);
\label{3-67}\\
&\ldots&\nonumber
\end{eqnarray}
\par Taking into account the boundary conditions for $X\to\pm\infty$,
the solution of the $\ell$-th equation in the hierarchy is
\begin{eqnarray}
S_{(\ell)}(X)&=&-\frac 1{2\kappa_\infty}\int_{-\infty}^\infty
{\rm d}\xi {\rm e}^{-\kappa_\infty|X-\xi|}\nonumber\\
&\phantom{=}&\phantom{-\frac{3-3}{2\kappa_\infty}\int_{-\infty}^\infty}
{\cal{F}}_\ell\left(\xi;S_{(1)},\ldots ,S_{(\ell-1)},v\right),
\label{3-68}\end{eqnarray}
which can be verified by direct substitution.

The large $\kappa_\infty$ limit of the first terms in the
perturbative expansion (\ref{3-62})  is
\begin{eqnarray}
S_{(0)}(X)+S_{(1)}(X)&=&1+\frac{\kappa_\infty}{2}\int_{-\infty}^\infty
{\rm d}\xi {\rm e}^{-\kappa_\infty|X-\xi|}v(\xi)\nonumber\\
%&\phantom{+}&\phantom{S_{(1)}}
&\to& 1+\frac 12 v(X),\label{3-69}
\end{eqnarray}
which coincides with the first terms in the perturbative expansion of the
"quasi-classical solution" (\ref{3-61}).
Several terms we checked also have this property.

Having found $S(X)$, we can now find the coordinate transformation
by a simple quadrature:
\begin{equation}
x=\int_0^X \frac{\rmd Y}{S(Y)}.
\label{3-quad}\end{equation}
In fact, this yields the inverse transformation, (\ref{3-4}).
Finally, the mass profile in the BDD equation is given in parametric form
by (\ref{3-quad}) and 
\begin{equation}
m(X)=m_\infty S(X).
\label{3-masp}\end{equation}
The concentration profile can now be readily recovered.

\subsection{Example}

We will illustrate the method by finding the specifications
for a filter with two narrow transmission resonances, centered 
at $E_1<E_2$, with the  high-energy resonance significantly 
narrower than the low-energy one\cite{BMMV97}. Apparently, 
this is difficult to achieve by the direct method\cite{ca-pc}.

We will start by constructing the Pad\'e approximant for the reflectance.
We will chose $q=1$ in (\ref{R_E}). This ensures that the potential
in the Schr\"odinger equation (\ref{3-1}) will be continuous.
We will try to keep $p$, the degree of the denominator in (\ref{R_E}),
ass small as possible. Choosing the reflectance to be equal
to zero at the energies $E_1$ and $E_2$, we find that we need at least $p=4$.
This means two double real zeros at $E_1$ and $E_2$. Thus, the denominator is
\begin{equation}
\calP_4(E)={\cal{A}}_2(E){\cal{B}}_2(E)=(1-E/E_1)^2(1-E/E_2)^2.\label{deno}
\end{equation}

There is some freedom in choosing the numerator, $\calQ_7(E)$.
Baring other constraints, it is convenient to note that $\calQ_7(E)$
has to have a pair of complex conjugate zeros in close vicinity to
each of transmittance maxima $E_1$ and $E_2$. We will choose
\begin{equation}
\calQ_7(E)=[{\cal{A}}_2(E)+E/F_1][{\cal{B}}_2(E)+E/F_2]\calQ_3(E).\label{num1}
\end{equation}
Here $F_1$ ($F_2$) must be much larger than $E_1$ ($E_2$), to ensure
that the zeros of the factors in the square brackets are close to
the corresponding zeros of the denumerator. Indeed, in these conditions,
the two poles near $E_1$ are given by
\begin{equation}
E \approx E_1\left(1 \pm \rmi\sqrt{E_1/F_1}\right),
\label{polic}\end{equation}
and a similar expression for the poles near $E_2$.
It remains to determine the positive polynomial $\calQ_3(E)$ in (\ref{num1}).

Since we want sharp transmittance maxima near $E_1$ and $E_2$, we
want that the background reflectance,
\begin{equation}
\calR_b(E)=1/Q_3(E),
\label{re-b} \end{equation}
will be close to 1 and relatively slowly varying in the range of
energies $0,\overline{E}$, with $\overline{E}>E_2$.
This can be achieved by choosing 
\begin{equation}
\calQ_3(E)=1+\frac{\delta}2
\left[1+\Bigl(\frac{2E}{\overline{E}}-1\Bigr)^3\right].
\label{q3-1} \end{equation}
Then, if  $\delta\ll 1$, the background reflectance is monotonically 
decreasing with $\calR_b(0)=1$ and $\calR_b(\overline{E})\approx 1-\delta$. 
It has a horizontal inflection point at $E=\overline{E}/2$ with 
$\calR_b(\overline{E})\approx 1-\delta/2$. 
 
Neglecting $\delta$ with respect to 1, the full widths at half maximum (FWHM) 
of the transmittance maxima are, respectively, $2\sqrt{E_1^3/F_1}$ and 
$2\sqrt{E_2^3/F_2}$. Thus, $F_1$ and $F_2$ can be obtained from the FWHM
of the corresponding resonance.

We chose $E_1=40meV$ and $E_2=100meV$, with FWHMs equal to $9meV$ and,
respectively, $3meV$ as initial data. Proceeding as explained in sections
\ref{pha}, \ref{rat}, \ref{bv-sol} we obtained an effective-mass profile
which yielded the same reflectance as the input one. This continuous profile
was digitized manually into 12 steps with lenghths which are integer numbers
of lattice constants and heighths which are a combination
of three concentrations, as described in Table 1. The $Al$ concentration
of the embedding alloy is 11.4\%.
In Fig. 1 we present
the transmittance calculated for the resulting configuration.
The digitization has introduced a few artefacts (the shoulder of the
low energy line and a low amplitude broad maxima, one of which is visible
in Fig. 1). The maxima were broadened and shifted a little from the design data.
Nevertheless, the ratio of FWHM of the
high/low energy maxima of transmittance is better than 2:1.

\section{Determination of chemical composition and dopant concentration
profiles of heterostructures with preset reflectance}\label{v}

In the section \ref{rat} we have shown how, given rational expressions
for the scattering data (SD), one can construct the piecewise continuous
potential $V(x)$, decaying exponentially at $+\infty$.
The one-dimensional Schr\"odinger with $V(x)$ has the given SD.
In section \ref{vmm} we mapped unitarily the one-dimensional Schr\"odinger
equation onto a family of BenDaniel and Duke type equations depending on
the function $X(x)$ which defines the transformation.
We also discussed in some detail the solution of the nonlinear
differential equation, which determines the effective mass profile 
in the case of an undoped heterostructure, with vanishing density of
conduction electrons. 

Now, we want to deal with the case when a nonzero density of conduction
electrons is present and the potentials, (\ref{S2}) in the SE, (\ref{S1}),
or in the BDD equation, (\ref{S3}), are self-consistent.
Let the external potential $U_{ext}(z)=0$ so that the potential
in the BDD equation is
\begin{equation}
U(z)=\calE_{cond}[c(z)]+\Phi_{sc}(z;m;U),
\label{4-1}\end{equation}
where we evidenced the functional dependence of $\Phi_{sc}(z;m;U)$ on
the effective mass profile $m(z)$ and the full potential energy $U(z)$.
In the SE case the effective mass is constant and $\Phi_{sc}$ depends only
on $U(z)$.
Inspection of (\ref{4-1}) suggests the idea that for a given full potential
$U(z)$, we can determine the chemical composition profile by moving the
selfconsistent potential to the left hand side of (\ref{4-1}).
We will see that in the SE case this is relatively easy to do
at a given operating temperature $T_o$. The situation is trickier in
the case of BDD dynamics, where we will present a perturbative approach
to solving the functional equation which replaces the nonlinear
differential equation (\ref{3-15}).

Let us start by considering the Hartree approximation for the self-consistent
electrostatic potential. Then, $\Phi_{sc}(z;m;U)$ is the solution of Poisson's
equation
\begin{equation}
\left[\varepsilon(z)\Phi_{sc}^\prime(z)\right]^\prime= 4\pi e\rho_{ch}(z).
\label{4-2}\end{equation}
Here $-e$ is the electron charge, $\varepsilon(z)$ is the dielectric constant, 
$\varepsilon(z)=\varepsilon[c(z)]$, and $\rho_{ch}(z)$ is the full charge
density, the difference between the ionized donor charge density and the 
electron one.
\begin{equation}
\rho_{ch}(z)=e\left[n_d(z)-n_{el}(z)\right].
\label{44-2a}\end{equation}
The density of donor dopant ions is made of a uniform
background density $n_b$ and the local variation of the 
density of donors in the heterostructure $n_\ell(z)$, which goes to
zero for large $|z|$:
\begin{equation}
n_d(z)=n_b + n_\ell(z).
\label{44-2b}\end{equation}
We assume that the doping and temperature are such that all the donors are
ionized and that the density of holes is negligible compared to the density
of donors. Then, we can neglect the valence bands. Otherwise, a multi-band 
treatment is needed. 

The electron dynamics is described either by the SE, (\ref{S1}), or by the BDD
equation, (\ref{S3}). The equilibrium electron density, $n_{el}(z)$, can be
calculated from the density of states for the corresponding equation,
(\ref{S1}) or (\ref{S3}),
\begin{equation}
n_{el}(z)=\int_{-\infty}^{+\infty}\frac{\rmd E\,\, \nu(z,E)}
{1+\rme^{\beta(E-\mu)}},
\label{4-3}\end{equation}
where $\beta$ is the inverse temperature (in energy units) and $\mu$ is
the chemical potential of the electrons. 
                      
A necessary condition for the stability of the system is that the full
charge is equal to zero,
\begin{equation}
\int_{-\infty}^{+\infty}\rmd z\rho_{ch}(z) =0.
\label{4-4}\end{equation}
In particular, the limiting value of electron density at infinity
equals the background ion density, $n_{el}(\pm\infty)=n_b$.
The limiting values of the potential at $\pm\infty$ will be equal only if
dipolar moment of the charge density is zero,
\begin{equation}
\int_{-\infty}^{+\infty}\rmd z\, z\, \rho_{ch}(z) =0.
\label{4-5}\end{equation}
Thus, $z^2\rho_{ch}(z)$ must go to zero at infinity and the solution
of (\ref{4-2}) is
\begin{equation}
\Phi_{sc}(z;m;U)=2\pi e\int_{-\infty}^{+\infty}\rmd u\rho_{ch}(u)
\left\vert\int_u^x\frac{\rmd v}{\varepsilon(v)}\right\vert. 
\label{4-6}\end{equation}
For position-independent $\varepsilon$, which we will consider in the case
of SE dynamics, (\ref{4-6}) becomes the well-known
\begin{equation}
\Phi_{sc}(z;U)=\frac{2\pi e}{\varepsilon}
\int_{-\infty}^{+\infty}\rmd u |z-u|\rho_{ch}(u).
\label{4-7}\end{equation}

We will consider separately the cases when the electron dynamics
is described by the Schr\"odinger equation, (\ref{S1}), and by the
BenDaniel and Duke equation, (\ref{S3}).

\subsection{Schr\"odinger's equation}\label{scSch}

To compute $\Phi_{sc}(z;U)$ using (\ref{4-7}), we need the electron density
of states in (\ref{4-3}). Since the transverse degrees of freedom
separate, (\ref{none}-\ref{0-s1}), we can integrate over the transverse
quasimomenta and express the three dimensional density of states $\nu_S(z,E)$
through the density of states $\nu_{0}(z,E)$ of the one-dimensional SE
(\ref{3-1}):
\begin{equation}
\nu_S(z,E)=\frac{m_e}{\pi\hbar^2}\int_{0}^{\infty}\rmd \eta \nu_{0}(z,E-\eta).
\label{4-9}\end{equation}
The one dimensional density of states $\nu_{0}(z,E)$ is proportional to
the imaginary part of the Green function $G_0(z,z^\prime;E)$ of the
one-dimensional SE (\ref{3-1})
\begin{equation}
\nu_{0}(z,E)=
-\frac 1\pi \lim_{\delta\downarrow 0}\Im\left[G_0(z,z;E+i\delta\right].
\label{4-10}\end{equation}
This allows a standardized treatment of the bound and continuum states.

For all $E$ with $\Im(E)\ne 0$, the Green function is the solution of the equation
\begin{equation}
\left[E+\frac{\hbar^2}{2m_e}\frac{\rmd^2}{\rmd z^2}-U(z)\right]
G_0(z,z^\prime;E)=\delta(z-z^\prime),
\label{4-11}\end{equation}
which is continuous at $z=z^\prime$ and goes to zero for $z\to\pm\infty$.
In terms of  the Jost functions, defined in Appendix \ref{appa-1},
the solution is
\begin{eqnarray}
G_0(z,z^\prime;E)&=&\frac{m_eT(k)}{\rmi k\hbar^2}
f_+(z;k)f_-(z^\prime;k); ~~z>z^\prime,\nonumber\\
&=&G(z^\prime,z;E);~~z<z^\prime.
\label{4-15}\end{eqnarray}
Here $k=\sqrt{2m_eE}/\hbar$ and $T(k)$ is the transmission coefficient.

Substituting (\ref{4-9}) and (\ref{4-10}) into (\ref{4-3}) we obtain
the electron density
\begin{equation}%
n_{el}(z)=\frac{m_e}{\pi\beta\hbar^2} \int_{-\infty}^{\infty}\rmd E
\ln\left[1+\rme^{\beta(\mu-E)}\right]
\nu_0(z,E).\label{4-14}
\end{equation}

Now, let us fix the {\it operating temperature} $T_o=1/\beta_o$.
The local dopant concentration profile, $n_\ell(z)$, will be chosen such
that the conditions (\ref{4-4}) and (\ref{4-5}) hold at $T_o$.
This can be done in many ways. For each  $n_\ell(z)$,  (\ref{4-1}) yields
a band offset profile
\begin{equation}
{\calE}_{cond}[c(z)]=U(z)-\Phi_{sc}(z;U).
\label{4-16}\end{equation}
If the values of the local doping profile $n_\ell(z)$ and the
concentration profile, $c(z)$, which can be determined from (\ref{4-16})
are physically achievable, a heterostructure with these specifications
can be manufactured. Assuming Schr\"odinger dynamics for the conduction
electrons, the selfconsistent potential at the operating temperature $T_o$
will be $U(z)$. At $T_o$, the SD and the bound states of the electrons in the
heterostructure will be those determined by $U(z)$.

Since the selfconsistent potential varies slowly with the temperature,
at temperatures which not too far from $T_o$, the selfconsistent potential
of the structure, which can be computed by standard means, will be close to
$U(z)$. An important result of scattering theory, see {\it e.g.}\cite{CS},
is that the SD and the bound states of the SE are continuous functionals of
the potential. Thus, the SD and bound states in an interval of temperatures
near $T_o$ will be close to those at $T_o$.

Now, a big advantage of the inverse method becomes apparent.
In the direct approach we give the composition profile $c(z)$
and the dopant profile $n_\ell(z)$, and have to compute the selfconsistent
potential. This is a rather time-consumming iterative process, which has to
be repeated after each modification of the chemical and dopant profiles.
Only after this are we able to obtain the scattering data and check
if they are desirable. In the inverse approach, we start from 
desirable SD and can obtain a whole class of specifications, parameterized
by the local dopant profile $n_\ell$, which yield the desired properties
exactly at $T_o$ and approximately in some neighborhood of $T_o$.

\subsection{BenDaniel and Duke's equation}\label{scBDD}

In section \ref{vmm} we have shown that to each monotonically
increasing on $(-\infty,\infty)$ function $X(x)$, one can associate
a unitary operator, $\hat{\sf{U}}_X$, which maps the SE onto a BDD type equation.
There we considered only undoped heterostructures without occupied
conduction electron states. 
Given the material relation between the band offset and the effective mass,
we have shown how to solve the nonlinear differential equation
which determines the transformation $X(x)$ and obtain the band offset
(and effective mass) profile which has the same SD as the initial SE.

As in section \ref{diff}, we assume that the effective mass' dependence
on the local chemical composition is invertible. Then, 
material relations exist in terms of the effective mass for the band offset,
(\ref{3-32}), and the {\it dielectric constant},
\begin{equation}
\varepsilon(z)=\epsilon[m(z)].
\label{4-21}\end{equation}

The electrostatic part of the selfconsistent potential, $\Phi_{sc}(z;m;U)$, 
is given by (\ref{4-6}) in terms of $\epsilon[m(z)]$ and the full charge
density $\rho_{ch}(z;m;U)$. We can now repeat the calculation of the electron
density of states like in section \ref{scSch}. The transverse degrees of
freedom separate again and, after performing the angular integral over
the transverse quasi-momenta

The electron density is given
again by (\ref{4-3}) where the density of states is
the integral over the transverse momenta of the one dimensional
density of states. S
etting $q_\perp^2=2m_\infty\eta/\hbar^2$ and
performing the angular integral over the transverse quasi-momenta
we obtain
\begin{equation}
\nu(z,E)=\frac{m_\infty}{\pi\hbar^2}\int_{0}^{\infty}\rmd \eta 
\nu_{\eta}(z,E-\eta),
\label{4-22}\end{equation}
where $\nu_{\eta}(z,E)$ is the density of states of the one-dimensional 
BDD equation (\ref{3-2}) with the potential $V_{BDD}$ replaced by
$V_\eta(z)=2m_\infty U_\eta(z)/\hbar^2$ where
\begin{equation}
U_\eta(z)=U_{BDD}(z)+ \eta \left[1-\frac{m_\infty}{m(z)}\right]
\label{4-22a}\end{equation}
The density of states for the one dimensional BDD equation can again be
expressed  in terms of its Green function, 
\begin{equation}%
\nu_{\eta}(z,E)=-\frac{1}{\pi}\lim_{\delta\downarrow 0}
\Im\left[G_{\eta}(z,z;E+\rmi \delta)\right].
\label{4-23}\end{equation}
For all $E$ with $\Im(E)\ne 0$, the Green function $G_{\eta}(z,z^\prime;E)$
is the solution of the equation
\begin{equation}%{eqnarray}
\left[E+\frac{\rmd}{\rmd z}\frac{\hbar^2}{2m(z)}\frac{\rmd}{\rmd z}
-U_\eta(z)\right] %\biggr\}
G_{\eta}(z,z^\prime;E) %\nonumber\\ &=&i
=\delta(z-z^\prime),
\label{4-24}\end{equation}%{eqnarray}
which is continuous for $z=z^\prime$ and goes to zero for $z\to\pm\infty$.
Like in the case of the SE, the one dimensional BDD Green function can be 
expressed in terms of the Jost solutions, $f_\pm(z^\prime;k;\eta)$ of the BDD 
equation (\ref{3-2}) with potential (\ref{4-22a}) 
which satisfy the boundary conditions (\ref{A3}).
Here, $k=\sqrt{2m_\infty E}/\hbar$. %
We will write here only the contact value needed in (\ref{4-23})
\begin{equation}
G_{\eta}(z,z;E)=\frac{m_\infty T(k)}{\rmi\hbar^2 k}
f_+(z;k;\eta)f_-(z;k;\eta). 
\label{4-27}\end{equation}

Substituting (\ref{4-22}) into (\ref{4-3}) and making the change of
variable $E\to E+\eta$ in the double integral, we obtain
\begin{equation}
n_{el}(z)=\frac{m_\infty}{\pi\hbar^2}
\int_{-\infty}^{+\infty}\rmd E\int_{0}^{\infty}\rmd \eta
\frac{\nu_{\eta}(z,E)}{1+\rme^{\beta(E+\eta-\mu)}}.
\label{4-28}\end{equation}

In the SE case, discussed above in subsection \ref{scSch},
$m(z)=m_\infty=m_e$ and $\nu_{\eta}(z,E)$ does not depend on $\eta$.
This allowed us to perform explicitly the integration with respect
to the energy of transverse motion, $\eta$, and obtain (\ref{4-14}).
Now, the mini-bands are not parabolic, and $\nu_{\eta}(z,E)$
depends on $\eta$. Nevertheless this dependence is weak for sufficiently
small $\eta$.
Let $T_o=1/\beta_o$ be the operating temperature.

In (\ref{4-28}), with exponentially small error, the main contribution 
to the $\eta$ integral comes from the interval
$(0,\mu-\underline{E}+CT_o)$, where $\underline{E}$ is
the bottom of the spectrum and the constant $C$ is several units, 
Considering only band offsets which are linear 
in the effective mass, (\ref{3-32}), the potential (\ref{4-1}) is
\begin{equation}
U_{BDD}(x)={B}[m(x)-m_\infty]+\Phi_{sc}(x;m;U).
\label{4-29}\end{equation}
so that
\begin{equation}
U_{\eta}(z)={B}[m(z)-m_\infty][1-\eta/{B}m(z)]+\Phi_{sc}(x;m;U).
\label{4-29e}\end{equation}
We see the condition for neglecting the nonparabolicity contribution to 
the electron density is that $(0,\mu-\underline{E}+CT_o)$  is much smaller
than the typical values of ${B}m(z)$ [800-1000meV for $Al_xGa_{1-x}As$,
see (\ref{3-B})].

After presenting the explicit recipe for computing the functional 
$\Phi_{sc}(z;m;U)$, let us apply the unitary transformation $\hat{\sf{U}}_X$,  
(\ref{3-5}), associated with the coordinate transformation (\ref{3-3})
to the one dimensional Schr\"odinger equation, (\ref{3-1}) 
with potential $V_S(z)$. We obtain the BDD-type equation (\ref{3-8}) 
with the potential (\ref{3-9}) and effective mass (\ref{3-10}).

The condition for the two equations to be identical is the coincidence of
the potentials (\ref{4-29}) and $U_X(x)=\hbar^2 W_X(x)/2m_\infty$, where 
$W_X(x)$ is given by (\ref{3-9}).
\begin{equation}
U_X(x)=U_{BDD}(x). 
\label{4-32}\end{equation}
This can be rewritten as
\begin{eqnarray}
X^{\prime\, 2}(x)-1-v[X(x)]
&-&\kappa_\infty^{-2}\left\{\frac{2X^{\prime\prime\prime}(x)}{[X^\prime(x)]^3}-
\frac{5[X^{\prime\prime}(x)]^2}{[X^\prime(x)]^4}\right\}
\nonumber\\ &=&-\phi_{sc}(x;X^\prime;v).
\label{4-30}\end{eqnarray}
Here, $\kappa_\infty$ is defined by (\ref{3-35}) and we introduced 
the potentials measured in units ${B}m_\infty$:
\begin{equation}
v(X)=\frac{U_S(X)}{{B}m_\infty};~~
\phi_{sc}(x;X^\prime;v)=\frac{\Phi_{sc}(x;m;U)}{{B}m_\infty}.
\label{4-31}\end{equation}
Taking into account (\ref{3-9}), (\ref{3-10}), (\ref{3-32}) and
(\ref{4-21}), $m(x)$ and $W_X(x)$ are functionals of the coordinate
transformation, $X^\prime$, and the potential, $V_S$,  in the
original Schr\"odinger equation, (\ref{3-1}). We reparameterized the
dependence of  $\Phi_{sc}(x;m;U)$ in terms of these functions, as it
appears in the second eq. (\ref{4-31}).

Setting the self consistency term, $\phi_{sc}(x;X^\prime;v)$, equal to zero
in (\ref{4-30}), we recover the differential equation (\ref{3-15})
in the linear case (\ref{3-32}). The presence of $\phi_{sc}(x;X^\prime;v)$
makes (\ref{4-30}) a functional equation.

We will attack this equation in a perturbative manner, as in
section \ref{map-ap}. Indeed, as we have seen there, in some interesting
cases $1/\kappa_\infty$ is of the order of the lattice spacing.
A "quasiclassical" approach to this is to neglect the term within braces
in the right hand side of (\ref{4-30}), leading to 
\begin{equation}
X^\prime(x)\approx \sqrt{1+v(X)-\phi_{sc}(x;X^\prime;v)}.
\label{4-qc}\end{equation}
Although this approximation seems reasonable, we will not pursue this further
with the same motivation as in section \ref{map-ap}.
Instead,  we will develop again the 
"small potential" approach, using the same motivation as 
in section \ref{map-ap}. Since $\phi_{sc}(x;X^\prime;v)$ is generally 
smaller and slower varying than $v[X(x)]$, the estimate of type (\ref{3-B})
will be better for $\phi$.

Before proceeding further, let us write the selfconsistency term in the right 
hand side of (\ref{4-30}) as a function of $X(x)$. As in section \ref{diff}, 
we can then reduce the order of the derivatives in the equation (\ref{4-30}) 
using the fact that it will no longer depends explicitly on $x$, but  
only through $X(x)$ and its derivatives. Redefining the unknown function, 
the selfconsistency correction's $\phi$ dependence on $x$ and the
two densities which appear in the definition of $\phi$ as
functions of $X(x)$,
\begin{eqnarray}
S[X(x)]&=&X^\prime(x); \label{4-def1}\\
\tilde{\phi}_{sc}[X(x);S;v]&=&\phi_{sc}(x;X^\prime;v); \label{4-deff}\\
\tilde{n}_{el}[X(x)]&=&n_{el}(x); ~~\tilde{n}_d[X(x)]=n_d(x).
\label{4-def2}\end{eqnarray}
and using (\ref{4-6}) and (\ref{4-31}), we obtain
\begin{equation}
\tilde{\phi}_{sc}(X;S;v)\hbox{\hskip -1pt}=\hbox{\hskip -1pt}
\left[
\int_{-\infty}^{X}\hbox{\hskip -2pt}\rmd z
\int_{z}^{X}\hbox{\hskip -2pt}\rmd t +
\int^{+\infty}_{X}\hbox{\hskip -2pt}
\rmd z\int^{z}_{X}\hbox{\hskip -2pt}\rmd t \right] g(z,t),
\label{4-33}\end{equation}
where the we introduced the notation
\begin{equation}
g(z,t)=
\frac{2\pi e^2 \left[\tilde{n}_d(z)-\tilde{n}_{el}(z)\right]}
{Bm_\infty S(z)S(t)\epsilon[m_\infty S^2(t)]}.
\label{4-34}\end{equation}
It remains to rewrite the density of states $\nu_\eta(x,E)$ as a 
function of $X(x)$. 
Let us apply the unitary transformation $\hat{\sf{U}}_X$ to the Schr\"odinger 
equation (\ref{3-1}) with the potential $V_S(z)$ replaced by 
\begin{equation}
V_{S\eta}(z)=V_S(z)+2m_\infty \eta\left[1-S^{-2}(z)\right]/\hbar^2.
\label{4-35-}\end{equation}
The function $S(z)$ was defined in (\ref{4-def1}).
The result of the transformation is the BDD equation (\ref{3-8}) with the 
potential replaced by
\begin{equation}
W_{X\eta}(x)=W_X(x)+2m_\infty \eta\left[1-S^{-2}[X(x)]\right]/\hbar^2.
\label{4-35}\end{equation}
Here $W_X$ is given by (\ref{3-9}). 
By (\ref{4-32}) and the definition of $S$, (\ref{4-def1}), we have that
$W_{X\eta}[X(x)]$ coincides with the potential in the equation
for the Green function (\ref{4-24}), $U_\eta(x)$. Thus, the Green function
$G_\eta(x,x^\prime,E)$ is the result of the unitary mapping with 
$\hat{\sf{U}}_X$ of the Green function of the Schr\"odinger equation 
(\ref{3-1}) with potential $V_S$ replaced by $V_{S\eta}$, (\ref{4-35-}). 
Finally, (\ref{4-23}) can be rewritten as
\begin{equation}
\nu_\eta(x,E)=S[X(x)]\nu_{0\eta}(X(x),E).
\label{4-37}\end{equation}
Here, $\nu_{0\eta}(z,E)$ is given by (\ref{4-10}) in terms of $G_{0\eta}(z,z;E)$. 
This is the Green function of the Schr\"odinger equation. It satisfies the 
equation (\ref{4-11}) with $m_e=m_\infty$ and the potential $U(z)$ replaced by 
$\hbar^2V_{S\eta}(z)/2m_\infty$.
Thus, substituting (\ref{4-37}) into (\ref{4-28}), we obtain 
the electron density as a function of $X(x)$ and a functional of
$S$ and $v$.

We can now rewrite (\ref{4-30}) in a form resembling (\ref{3-63}):
\begin{equation}
S^{\prime\prime}=3S^{\prime\,\, 2}/2S\, +\,
\kappa_\infty^2S\left[S^2 -1-v+\tilde{\phi}\right]/2\, ,
\label{4-38}\end{equation}
where we omitted the arguments of the functions.

The small potential perturbative approach begins with introducing 
the formal small parameter $\alpha$, setting
\begin{equation}
v\to\alpha v;~~\tilde{\phi}\to \alpha\tilde{\phi}
\label{4-alfa}\end{equation}
Then we expand again $S$ into the series (\ref{3-62}) in powers of $\alpha$.
We substitute (\ref{3-62}) and (\ref{4-alfa}) into (\ref{4-38}) and
expand it into a power series in $\alpha$. After equating the terms with
the same power of $\alpha$ and setting $\alpha=1$, we obtain
the "small potential" expansion.      

As in section \ref{map-ap}, the zero-th order coincides with a uniform 
structure. The  equation $S_{(0)}^{\prime\prime}(X)=0$, with the
boundary condition $S_{(0)}(\pm\infty)\to 1$, leads to
\begin{equation}
S_{(0)}(X)=1.
\label{4-39}\end{equation}
The problem with the rest of the series is that although the Green
function $G_0$ can be expanded in %(\ref{4-24}), in 
terms of the potential:
\begin{equation}
G_0=G^{(0})+G^{(0})UG^{(0})+\ldots ,
\end{equation}
where $G^{(0})(x,y;E)=\rme^{\rmi k|x-y|}/2\rmi k$ is the free particle
Green function, this expansion assumes that the potential is small 
compared to the kinetic energy.
This is certainly wrong at the low energies we are interested in.

We will note instead that the nonparabolicity correction to $V_S$ 
(the second term in (\ref{4-35-})) is small. To first order in 
the expansion of $S$ in powers of $\alpha$, using the estimate
(\ref{4-qc}), we obtain
\begin{equation}
U_{S\eta}\approx U_S+[U_S-\Phi_{sc}]\eta/Bm_\infty.
\label{4-40}\end{equation}
Here, $U_{S\eta}=\hbar^2V_{S\eta}/2m_\infty$ and $U_S=\hbar^2V_S/2m_\infty$ 
are, respectively, the potential (\ref{4-35-}) and the reconstructed potential 
in the Schr\"odinger equation measured in energy units. We omitted the 
arguments of the functions. If $\mu-\underline{E}+CT_o$ is much smaller
than $Bm_\infty$, then the term with square brackets in (\ref{4-40}) is
small compared to the first one.

Neglecting it, %we obtain that
the self consistency term coincides
in the first approximation with the one computed 
in section \ref{scSch} for the Schr\"odinger dynamics with mass
$m_\infty$ and dielectric constant $\epsilon(m_\infty)$:
\begin{equation}
\tilde{\phi}(X;1;v)=
\Phi_{sc}(X;U_S)/Bm_\infty.
\label{4-41}\end{equation} 

Now, we can solve for the first order correction $S_1(X)$:
\begin{equation}%
S_1(X)=\frac 1{4\kappa_\infty}\int_{-\infty}^\infty
{\rm d}\xi {\rm e}^{-\kappa_\infty|X-\xi|}
[v(\xi)-\tilde{\phi}(\xi)],
\label{4-x8}\end{equation}%{eqnarray}
with $\tilde{\phi}$ given by (\ref{4-41}). If the integrand is
slowly varying on the scale $1/\kappa_\infty$, then 
we recover the perturbative expansion of "quasiclassical" 
formula (\ref{4-qc}).
\begin{equation}%
2 S_1(X)=
v(X)-\tilde{\phi}(X;1;v).
\label{4-x9}\end{equation}

Finally, we find the transformation as at the end of
section \ref{map-ap}.
\begin{equation}
X(x)=x+\int_0^x \rmd y\, S_1(y),
\label{4-quad}\end{equation}
where we inverted the implicit relation (\ref{3-quad}) to first order.
The mass (and concentration) profile can now be recovered.
The result will be reasonable if the electron (and dopant) density
is not too large. A more detailed account will be published elsewhere\cite{prep}.

\section{Concluding remarks}\label{concluding}

We have attained our stated purpose, constructing
heterostructure chemical composition and dopant profiles such
that the electron reflectance of the heterostructure is preset,
by combining  three inverse techniques:
\begin{itemize}
\item[{\bf 1.}] Reconstruction of the phases of the scattering data
from their absolute values. We did it by Pad\'e discretization
of the dispersion relations.
\item[{\bf 2.}] Reconstruction of the potential in the one dimensional
Schr\"odinger equation from the scattering data. We used the
Kay-Moses-Sabatier approach, since the reconstructed scattering data
were rational functions of $k$.
\item[{\bf 3.}] Determination of the chemical concentration and dopant
ion concentration profiles which yield a given self consistent potential
for the Schr\"odinger equation or an effective mass profile and
a self consistent potential for the BenDaniel and Duke equation.
We simply reversed the definition of the self consistent potential in the
Schr\"odinger case. In the case of BenDaniel and Duke dynamics we
sketched a perturbative approach to solving the functional
equation obtained by the variable mass mapping of the Schr\"odinger
into the BenDaniel and Duke one -- the method used for solving
the non self consistent inverse problem for the BDD equation.
\end{itemize}

In section \ref{pha} we used Pad\'e approximation for representing
the reflectance and reconstructed the phases by discrete dispersion relations.
The number of parameters is determined by two factors:
the number of poles of the approximants for the transmittance and 
the reflectance ($p+q+2$) and the degree of smoothness of the
potential. The latter is determined by the rate of falloff at infinity
of the reflection coefficients. If $R_\pm(k)\sim k^{-q-2}$, then  the $q$-th 
derivative of the potential is piecewise continuous. The accuracy of
representation of the scattering data increases with increasing $p$.
So does the complexity of the calculations.

The reference solutions are in some sense maximally non-symmetric. This
can be advantageous in optimizing non linear optical response functions,
which vanish for symmetric potentials.

We have shown how to construct chemical profiles for compositionally graded
heterostructures in lattice matched systems, like the systems $Ga_{1-x}Al_xAs$ or 
$In_{1-x-y}Ga_xAl_y$ matched on $InP$, such that in the effective mass
approximation (with self-consistent potential) the conduction electron
reflectance (at a preset operating temperature) is given by the designer.
In the inverse approach, the construction of self-consistent potentials
can be made at little extra cost. Indeed, in the brute-force direct approach
one examines a number of possible configurations and, for each configuration,
one has to construct iteratively the self-consistent potential and
{\it only then} one can obtain the scattering data for the configuration. 

In the inverse approach for the SE, once the reference potential is
constructed from the synthesized scattering data, one can choose some doping
profile and determine directly the compositional profile which yields the
desired reflectance. The situation is somewhat more complicated in the case of
the BDD equation. Here, we start again from the reference potential. 
Then, solving a nonlinear differential equation, we obtained the concentration 
(mass) profile corresponding to a non-selfconsistent potential. This is 
valid if the conduction electron density is negligible. 

In the case of non negligible density of conduction electrons, a more
complicated functional equation must be solved. The main difficulties
are generated by the non-parabolic character of the minibands in the case of 
the BenDaniel and Duke's equation. Perturbative approaches were presented 
in section \ref{scBDD}. A hybrid approach which looks promising is 
to search by brute force methods near the perturbative solution in the 
BDD case, looks promising.

We examined only  Hartree self consistent potentials. It is obvious that
the incorporation of exchange-correlation corrections which depend
only on the local electronic density can be done at little extra cost.

We will mention without details another case in which the calculation of
the selfconsistent solution is much simplified. When the electron plasma is
thin and hot everywhere,
$e^2\beta\left[n_{el}\right]^{1/3}\ll 1$ and
$\hbar^2\beta\left[n_{el}\right]^{2/3}/m\ll 1$,
we may use the
(quasi)classical expression for the electron density:
\begin{equation}
n_{el}(z;m;U)=n_b\left[m(z)/m_\infty\right]^{3/2}\rme^{-\beta U(z)}.
\end{equation}

Throughout this paper we have considered the heterostructure embedded
in a homogeneous alloy and did not consider the biased case. The methods
described here can be generalized to the case of different asymptotic
compositions at $\pm\infty$ and/or presence of bias. This will be presented
elsewhere\cite{next}.

We have examined throughout this paper only continuous potentials.
If these are digitized using standard methods, the low energy dynamics 
and scattering will not be much affected.

If a more precise, multi-band description is needed, it should again
be profitable to explore by brute-force standard methods the configurations
which are close to the ones generated by our one-band approach. Perturbative
calculations near this point could also be prove useful.

Finally, let us summarize the advantages of the inverse approach over the
direct, brute force, one:
\begin{itemize}
\item{} In practice, the direct method is restricted to optimizing over small sets 
of parameters describing the structural data.
\item{} The inverse approach allows the discovery of {\it new} promising configurations 
that can be subsequently optimized using perturbative and/or traditional (brute force)
techniques.
\item{} Obtaining chemical composition and dopant concentration profiles and 
corresponding to given self-consistent potentials for the Schr\"odinger dynamics 
is rather inexpensive. Although the effort is significantly larger in the 
case of BenDaniel and Duke dynamics, good starting configurations
can be obtained perturbatively.
\end{itemize}

\appendix

\section{Scattering and inverse scattering for the one dimensional 
Schr\"odinger equation}\label{appA}

For the reader's convenience we will briefly review scattering and
inverse scattering theory for the one-dimensional Schr\"odinger
equation. We will use the Faddeev-Marchenko\cite{faddeev,M55,marche}
approach. A detailed account may be found in the book by Chadan and
Sabatier\cite{CS}.

Setting the in-plane quasimomentum $\mbq_\perp=0$ 
brings the one-dimensional Schr\"odinger equation (\ref{0-s1}) to the form 
\begin{equation}
\psi^{\prime\prime}(z)+\left[k^2-V(z)\right]\psi(z)=0.
\label{A1}\end{equation}
We will consider only potentials which are piecewise continuous and fall
sufficiently fast at infinity so that
\begin{equation}
\int_{-\infty}^{\infty}\rmd z (1+z^2)|V(z)|<\infty.
\label{c2}\end{equation}
Then, (\ref{A1}) may have only a finite number of bound states
with negative energy. For $E>0$ the spectrum is absolutely continuous
and doubly degenerate. 

\subsection{Jost functions, scattering data and their properties}\label{appa-1}
The scattering is best described in terms of the Jost solutions
of (\ref{A1}), $f_\pm(z;k)$, 
which for $k=0$ behave like outgoing waves near $\pm\infty$:
\begin{eqnarray}
\lim_{z\to +\infty}&f&_+(z;k)\rme^{-\rmi kz}=1;\nonumber\\
\lim_{z\to -\infty}&f&_-(z;k)\rme^{\rmi kz}=1.\label{A3}
\end{eqnarray}
Let us summarize some properties of the Jost functions (JF) which are
relevant for the scattering problem:

\begin{itemize}\label{jostf}
\item[{\it a.}] For real $k$ the JF are continuous in $z$. $f_+(z;\pm k)$ 
are a pair of linearly independent solutions of (\ref{A1}). The same
holds for $f_-(z;\pm k)$. 
\item[{\it b.}] The JF can be continued analytically from the positive half-axis ($k>0$) 
to the upper complex half-plane ${\rm Im}(k)>0$. Here, the JF are
{\it analytic} in $k$ with values which are continuous functions of $z$. 
They have no zeros in $z$ for ${\rm Im}(k)>0$. 
\item[{\it c.}] For large complex $|k|$ in the upper half-plane, the JF behave like 
outgoing waves for all real $z$:
\begin{equation}
\rme^{\mp\rmi kz}f_\pm(z;k)=1+{\cal{O}}(k^{-1});~~ |k|\to\infty.
\label{A4}\end{equation}
\end{itemize}

The above propositions can be readily proved using the integral equation 
of Volterra type which is satisfied by the JF:
\begin{equation}
f_\pm(z;k)=\rme^{\pm\rmi kz} +\int_z^{\pm\infty}\rmd y
\frac{\sin k(y-z)}k\, V(y)f_\pm(y;k). \label{volt}
\end{equation}

Now, we have two pairs of linearly independent solutions of (\ref{A1}),
$f_\pm(z;\pm k)$. One of each pair behaves like an outgoing/ingoing
near the corresponding infinity. Since (\ref{A1}) can have only two
linearly independent solutions, the outgoing wave JF  can be expressed
in terms of the ingoing wave ones:
\begin{eqnarray}
T(k)f_+(z;k)= f_-(z;-k)+R_-(k)f_-(z;k);\label{6p}\\
T(k)f_-(z;k)= f_+(z;-k)+R_+(k)f_+(z;k).\label{6m}
\end{eqnarray}
Instead of seeking the asymptotic behaviors, the transmission, $T(k)$ and 
the reflection coefficients to the right/left, $R_\pm(k)$, can be expressed 
in terms of Wronskian determinants of the JF:
\begin{eqnarray}
T(k)&=&\frac{2\rmi k}{W[f_+(z;k),f_-(z;k)]};\label{tr}\\
R_\pm(k)&=&\frac{W[f_-(z;\pm k),f_+(z;\mp k)]}{W[f_+(z;k),f_-(z;k)]};\label{rpm}
\end{eqnarray}
where we use the notation:
\begin{equation} 
W[f(z),g(z)]=f^\prime(z) g(z)-f(z)g^\prime(z),
\end{equation}
for the Wronskian of the functions $f(z)$ and $g(z)$. We remind the reader 
that the Wronskian of any two solutions of (\ref{A1}) does not depend on 
$z$ and is equal to zero if and only if the solutions are linearly dependent.

For real $k$ the scattering coefficients $T(k)$ and $R_\pm(k)$ 
satisfy the following relations:
\begin{eqnarray}
R_+(k)T(-k)+R_-(-k)T(k)&=&0,\label{7a}\\
T(k)T(-k)+R_\pm(k)R_\pm(-k)&=&1,\label{7b}
\end{eqnarray}
which express the unitarity of the $S$-matrix.
Reality of the potential implies also that
\begin{equation}
T(k)=[T(-k)]^*;~~~~R_\pm(k)=[R_\pm(-k)]^*.\label{8}
\end{equation}
Generically, $T(0)=0$ and $R\pm(0)=-1$ (if there are no "zero energy bound 
states" --- bounded, but not square integrable, solutions of the 
Schr\"odinger equation for $k=0$).
From (\ref{tr}) and the remark {\it c.} above we see that $T^{-1}(k)$ can be 
continued analytically to the upper half-plane, Im$(k)>0$. Its zeros, if 
present, are the only possible (simple pole) singularities of $T(k)$
in the upper half plane. At such a zero, the two JF $f_\pm(z;k)$ are not 
linearly independent and decay exponentially  for $z\to\pm\infty$. Thus, 
the poles of $T(k)$ in the upper half plane can occur only for 
$k=\rmi\lambda_j$, where $\hbar\lambda_j=\sqrt{-2m_eE_j}>0$ and $E_j$ are 
the energies of the bound states of (\ref{A1}). The corresponding
eigenfunctions $\psi_j(z)$ are real and are normalized by 
$\int_{-\infty}^{+\infty}\rmd x |\psi_j(x)|^2=1$. The two JF are proportional
to the bound-state wave function:
\begin{equation}
f_\pm(z;\rmi\lambda_j)=C_j^\pm\psi_j(z),\label{b-state}
\end{equation}
where $C_j^\pm$ are real constants.

The asymptotic expansion of $T(k)$ near the 
bound-state pole $k=\rmi \lambda_j$ is 
\begin{equation}
T(k)\approx \frac {\rmi }{C_j^+C_j^-}\frac 1{k-\lambda_j}
+\calO(1),\label{b-res}
\end{equation}
where $C_j^\pm$ are the constants in (\ref{b-state}).
From here we can see that the bound-state poles of $T(k)$ must be simple.
Otherwise, the product of the normalization constants is zero.

Generally speaking, the domains of analiticity of  
$R_\pm(k)$ will be smaller. 
If the potential $V(z)$ is zero on a half-axis and there are no bound states, then 
the corresponding reflection coefficient is analytic in the upper half-plane.
Let us check this for the reflection to the left coefficient, $R_-(k)$, and 
potentials which vanish for  $z<0$.  The denominator in (\ref{rpm}) is
analytic and has no zeros since there are no bound states. In the numerator,
$f_+(z;k)$ is always analytic and $f_-(z;-k)=\rme^{\rmi kz}$ for $z<0$. The
other case can be dealt with in a similar manner.

From (\ref{7a} - \ref{8}) we obtain relations between the analytic 
continuations of the scattering data, which are valid whenever
the arguments of the functions are within the domain of analiticity:
\begin{eqnarray}
R_+(k)T(-k)+R_-(-k)T(k)=0;\label{7ac}\\
T(k)T(-k)+R_\pm(k)R_\pm(-k)=1;\label{7bc}\\
{[T(k)]}^{*} = T(-k^*); ~~~
{[R_\pm(k)]}^{*} = R_\pm(-k^*).\label{8c}
\end{eqnarray}

For positive energies, $E=\hbar^2k^2/2m_e$, (\ref{7b}-\ref{8}) imply
that the sum of the transmittance of the heterostructure and its reflectance, 
\begin{equation}
\calT(E)+\calR(E)=
\left\vert T(k)\right\vert^2+\left\vert R_\pm(k)\right\vert^2=1
\end{equation}
equals unity. 

\subsection{Inverse scattering. The Marchenko equation.}\label{i-M}

If the potential in the Schr\"odinger equation (\ref{A1}) is known, then,
solving (\ref{A1}) and using (\ref{6p}-\ref{6m}) one can obtain
the scattering data (\ref{sd}). 
We want to explore the possibility of recovering the potential in the
Schr\"odinger equation (\ref{A1}) from the SD (\ref{sd}).

Let 
\begin{equation}
F_\pm(z;k)=\rme^{\mp\rmi kz}f_\pm(z;k).\label{Fpm}
\end{equation} 
The functions $F_\pm(z;k)$ are
analytic in the upper half-plane. As a function of $k$, $F_\pm(z;k)-1$ decays 
at infinity no slower than $k^{-1}$, (\ref{A4}). This means that the Fourier
transformation with respect to $k$ will exist at least in the $L^2$ sense.
We define the transformation kernels
of the Schr\"odinger equation by the Fourier transforms
\begin{eqnarray}
K_+(x;y)&=&\frac 1{2\pi}\int_{-\infty}^{+\infty}\rmd k 
\rme^{-\rmi k(y-x)}\left[F_+(x;k)-1\right],\label{A9-}\\
K_-(x;y)&=&\frac 1{2\pi}\int_{-\infty}^{+\infty}\rmd k 
\rme^{\rmi k(y-x)}\left[F_-(x;k)-1\right].\label{A9}
\end{eqnarray}
Closing the integration contour in the upper half-plane, we see that 
\begin{equation}%{eqnarray}
K_+(x,y)=0 \hbox{\rm ~for~} x>y;~~~\\ 
K_-(x,y)=0 \hbox{\rm ~for~} x<y.\label{Kpm}
\end{equation}%{eqnarray} 
The inverse Fourier transformations are:
\begin{eqnarray}
F_+(x;k)&=&1+\int_x^{+\infty}\rmd u \rme^{\rmi k(u-x)}K_+(x;u),\label{A10-}\\
F_{-}(x;k)&=&1+\int^x_{-\infty}\rmd u \rme^{\rmi k(x-u)}K_+(x;u).\label{A10}
\end{eqnarray} 

Taking now the Fourier transformations of the Volterra equations, (\ref{volt}),
satisfied by the JF yields 
\begin{eqnarray}
K_+(x;y)=
\frac 1{2}\int_{\frac{x+y}{2}}^{+\infty}\rmd s
\bigg[&V&(s)+\nonumber\\ 
+ 2\int_0^{\frac{y-x}{2}}\rmd t V(s-t)&K&_+(s-t;s+t)\bigg].\label{A11-}\\
K_-(x;y)=
\frac 1{2}\int^{\frac{x+y}{2}}_{-\infty}\rmd s
\bigg[&V&(s)+\nonumber\\ +
2\int^0_{\frac{y-x}{2}}\rmd t V(s-t)&K&_-(s-t;s+t)\bigg].
\label{A11}
\end{eqnarray} 
Finally, taking the limit $y\to x\pm 0$ in (\ref{A11-}-\ref{A11}) we
obtain a simple relation between $K_\pm$ and the potential $V(x)$:
\begin{eqnarray}
\lim_{y\downarrow x+0}K_+(x;y)&=&
\frac 1{2}\int_x^{+\infty} \rmd s V(s)\label{VK-}\\
\lim_{y\uparrow x-0}K_-(x;y)&=&
\frac 1{2}\int^x_{-\infty} \rmd s V(s).\label{VK}
\end{eqnarray} 
Thus, if we can construct $K_+$ (or $K_-$) from the scattering data,
we may use one of the relations (\ref{VK-}-\ref{VK}) to recover the 
potential by simple differentiation. 

To make this connection, we will use the analytic properties of the scattering
data. Applying Cauchy's formula to the analytic function $F_-(x;k)-1$, which
decays at infinity no slower than $k^{-1}$, we have
\begin{equation}
F_-(x;k)-1=\lim_{\delta\downarrow +0}
\frac 1{2\pi\rmi}\int_{-\infty}^{-\infty}\rmd\kappa 
\frac{F_-(x;\kappa)-1}{\kappa -k-\rmi\delta}.  \label{A12}
\end{equation}
Changing $\kappa\to -\kappa$ in (\ref{A12}) and  using (\ref{6m}) yields 
\begin{eqnarray}
F_-&(&x;k)=1+\lim_{\delta\downarrow 0}
\frac 1{2\pi\rmi}\int_{-\infty}^{-\infty}\rmd\kappa % \Big [ 
\frac{1-T(\kappa ) F_+(x;\kappa)}
{\kappa +k+\rmi\delta}+\nonumber\\
&\phantom{(}&+\lim_{\delta\downarrow 0}
\frac 1{2\pi\rmi}\int_{-\infty}^{-\infty}\rmd\kappa % \Big [ 
\frac{R_-(\kappa ) F_-(x;\kappa)\rme^{-2\rmi\kappa x}}
{\kappa +k+\rmi\delta}.  %\Big ].
\label{A13} \end{eqnarray}

In the first integral on the right hand side of (\ref{A13}) we may 
close the contour in the upper half-plane. 
If there are no bound states, the integrand is 
analytic in the upper half-plane, and the integral is equal to zero.
If bound states are present, the first integral in (\ref{A13}) is
equal to
\begin{equation}
I_1=-\rmi\sum_j
\frac{\left(C_j^-\right)^2\rme^{2\lambda_jx}F_-(x;\rmi\lambda_j)}
{k+\rmi\lambda_j},\label{int1}
\end{equation}  
where we used (\ref{b-state}) to replace $F_+(x;\rmi\lambda_j)$
by $F_-(x;\rmi\lambda_j)$.

Taking the Fourier transform of (\ref{A13}), using (\ref{A9}) and 
the definition, (\ref{A10}), of $K_-(x;y)$,  yields the 
{\it Marchenko equation} for $K_-$:
\begin{equation}
K_-(x;y)+M_-(x+y)\hbox{\hskip -1pt}+\hbox{\hskip -1pt}
\int_{-\infty}^x\hbox{\hskip -2pt}\rmd s M_-(y+s)K_-(x;s)=0.\label{A14}
\end{equation}
Here, the Marchenko kernel, $M_-(u)$, is the sum of the 
Fourier transform of reflection to the left coefficient $R_-(k)$ 
and the contribution from the bound states, (\ref{int1}):
\begin{equation}
M_-(u)={1\over{2\pi}}\int_{-\infty}^{+\infty}\rmd\kappa 
\rme^{-\rmi\kappa u} R_-(\kappa)
+\sum_{j}\left(C_j^-\right)^{-2}\rme^{\lambda_ju}.\label{A15}
\end{equation}
A similar calculation yields the Marchenko equation for $K_+$:
\begin{equation}
K_+(x;y)+M_+(x+y)\hbox{\hskip -1.5pt}+\hbox{\hskip -1.5pt}
\int^{+\infty}_x\hbox{\hskip -4pt}\rmd s M_+(y+s)K_+(x;s)=0.\label{A14plus}
\end{equation}
The Marchenko kernel $M_+(u)$ is given by an expression similar to
(\ref{A15}), with all the ${-}$ signs changed into ${+}$ and 
$\lambda_j$ replaced by $-\lambda_j$.

The Marchenko equation links {\it directly} the SD to the
transformation kernels $K_\pm$ {\it bypassing the wave functions}.
After solving (\ref{A14}), the potential $V(x)$ is recovered from
(\ref{VK})
\begin{equation}
V(x)=2\frac{\rmd}{\rmd x} K_-(x;x-0).\label{A-VK}
\end{equation}
Let us remark that the first variable appears only as a parameter in 
(\ref{A14}) and (\ref{A14plus}).

Deift and Trubowitz\cite{DT79} have given necessary and sufficient conditions
for a one-to-one correspondence between a set of SD
$\{T(k), R_\pm(k)\}$ which has no bound states, and a potential 
$V(z)$ satisfying (\ref{c2}) obtained by solving Marchenko's equation, 
(\ref{A14}). 
\begin{itemize}\label{Deift}
\item[{\it 1.}] $T(k)$ and $R_\pm(k)$ satisfy (\ref{7a}-\ref{8}) on the real 
axis.
\item[{\it 2.}] $T(k)$ is analytic in the upper half-plane and continuous
up to the real axis. 
\item[{\it 3.}] $T(k)=1+{\calO}(|k|^{-1}), ~\Im(k)\ge 0, $ and 
$R_\pm(k)={\calO}(|k|^{-1})$, for real $k$ as $|k|\to\infty$. 
\item[{\it 4.}] $T(k)$ has no zeros on the real axis, excepting possibly
a simple one for $k=0$. In the latter case, 
\begin{equation}
1+R_\pm(0)=0.\label{R-0}
\end{equation}
\item[{\it 5.}] The Marchenko kernels $M_\pm(x)$ are absolutely continuous 
and for any given $a>0$ there exists $c(a)>0$ such that 
\begin{equation}
\int_{-\infty}^{+\infty}
\rmd x\theta[\pm (x-a)] (1+x^2) |M^\prime_\pm(x)| < c(a).\label{D-T}
\end{equation}
\end{itemize}

The problem becomes a little trickier in the presence of bound states.
Then, the Marchenko kernels $M_\pm(u)$, (\ref{A15}), depend now not only on 
the energy of the bound state, which is given by the corresponding
pole of $T(k)$, but also on the constants $C_j^\pm$, (\ref{b-state}). 
The scattering data contain information only on the product $C_j^-C_j^+$,  
which can be recovered from the residue of $T(k)$, (\ref{b-res}).
For each bound state % $j$
we can choose
one of the parameters $C_j^\pm$ arbitrarily. The other is fixed by
the (analytically continued) scattering data. 
Thus,  assuming full knowledge of the scattering data, in the case
when there are $n\ge 1$ bound states the solution of 
the inverse problem is not unique. There is a $n$-parameter
family of potentials which correspond to the same scattering data.

From a physical point of view, one cannot recover the full information 
on the bound states in scattering experiments, which study 
only the behavior of the solutions at large distances, where
the relevant information on the bound states is exponentially vanishing.

The numerical solution of the Marchenko equation for potential reconstruction 
is expensive from the computational point of view. To find a value 
for $V(x)$ one has to solve (\ref{A14}) with high enough precision for
the subsequent numerical differentiation. A lot of useless data is generated
in the process, since we need only $\lim_{y\uparrow x}K_-(x;y)$ for using
(\ref{A-VK}) (or $\lim_{y\downarrow x}K_+(x;y)$). 
We will solve the Marchenko equation in the manner explained in
section \ref{rat}, which is closer to the way we solve the
phase reconstruction problem in section \ref{pha}. 
The resource management compares rather favorably to that of the
codes\cite{FJ90,FJ92} which have been written for the
direct solution.

\vbox{\vskip 2em}

{\bf Acknowledgments:} We thank Tom Gaylord and Elias Glytsis for introducing us 
to the fascinating subject of heterostructure design and for fruitful
discussions. Illuminating discussions with Roger Balian, 
Pierre Sabatier and Giorgio Mantica are gratefully acknowledged. 
% Daniel Vrinceanu's assistance with computations has been invaluable. 
G.A.M. thanks Alfred Msezane and Carlos Handy for support and hospitality 
at C.T.S.P.S., Clark-Atlanta University, where significant parts of the
work presented here have been done. 

%$^{*}$ E-mail: mezin@alpha1.infim.ro

%\newpage

\begin{table}%[h]
\caption{12 layer digitized Al$_c$Ga$_{1-c}$As filter. The
Al concentrations are $c_1=0.05714,~c_2=2c_1,~c_3=4c_1;$ the bulk
Al concentration is $c_2=11.4\%.$\hfill}
\label{table1}
\begin{tabular}{cccc}
Layer&Width&Width&Al\cr 
\# &(Atomic layers) &(nm) &concentration \cr
\hline
1&6&1.696&$c_1+c_2+c_3$\\ \hline
2&9&2.543&0\\ \hline
3&18&5.088&$c_3$\\ \hline
4&5&1.413&$c_2$\\  \hline
5&10&2.827&$c_1+c_2$\\ \hline
6&10&2.827&$c_1$\\   \hline
7&14&3.957&$c_1+c_2$\\  \hline
8&14&3.957&$c_1$\\  \hline
9&13&3.675&$c_1+c_2$\\  \hline
10&13&3.675&$c_1$\\ \hline
11&14&3.957&$c_1+c_2$\\ \hline
12&11&3.109&$c1$\\ \hline
\end{tabular}
\end{table}

%\vbox{\vskip 1.0truecm}

%\newpage

%\widetext
%\protect{\vbox{\vskip 2.0truecm}}

\begin{figure}[h]
\includegraphics{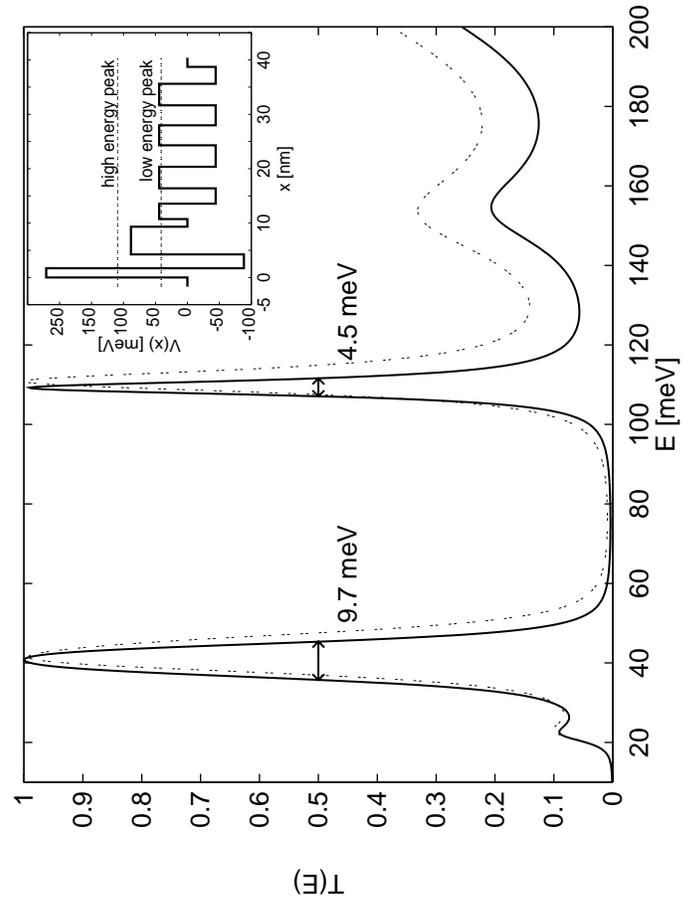}
\caption{Energy dependence of the transmittance of 12 layer digitized
$Al_cGa_{1-c}As$ filter: continuous line --- Eq.(\protect{\ref{3-2}});
dotted line --- constant mass approximation Eq.(\protect{\ref{3-1}})
with $m_0=m_\infty$. Insert: potential energy profile.\label{fig1}}
\end{figure}

\end{multicols}

\begin{thebibliography}{99}
\bibitem{confluence} We can regard the eventual double real zeros
of $\calP_p(E)$ at the transparency energies as confluences of complex
conjugate zeros ${\frak p}_{j_1}={\frak p}_{j_2}^*$. Then, in (\ref{p2}), 
the corresponding zeros satisfy $p_{j_1}=-p_{j_2}^*$.
\bibitem{smoo3} This smoothness condition ensures that the operator
domains of the initial Schr\"odinger and the resulting BDD type
Hamiltonians coincide. In plainer words, if $X(x)$ is less smooth, 
then the equation satisfied by the transformed function 
$\hat{\sf{U}}_Xf$ will be singular.
\end{thebibliography}
\end{document}